# Library Catalog Analysis and Library Holdings Counts: origins, methodological issues and application to the field of Informetrics

Daniel Torres-Salinas[1]
*Universidad de Granada, Dpto Información y Comunicación, EC3metrics SL*
Wenceslao Arroyo-Machado
*Universidad de Granada, Dpto Información y Comunicación*

**Abstract**

In 2009, Torres-Salinas & Moed proposed the use of library catalogs to analyze the impact and dissemination of academic books in different ways. Library Catalog Analysis (LCA) can be defined as the application of bibliometric techniques to a set of online library catalogs in order to describe quantitatively a scientific-scholarly field on the basis of published book titles. The aim of the present chapter is to conduct an in-depth analysis of major scientific contributions since the birth of LCA in order to determine the state of the art of this research topic. Hence, our specific objectives are: 1) to discuss the original purposes of library holdings 2) to present correlations between library holdings and altmetrics indicators and interpret their feasible meanings 3) to analyze the principal sources of information 4) to use WorldCat Identities to identify the principal authors and works in the field of Informetrics.

# 1. Introduction

In the Social Sciences and the Humanities, the evaluation of scientific activity and, especially, of the academic book, has been an unresolved issue because evaluation in bibliometry has, until quite recently, been monopolized by citation indexes and the Thomson Reuters databases (Nederhof, 2006). Hence, even though the vast majority of research studies demonstrate the importance of books in scientific communication (Archambault et al., 2006; Hicks, 1999; Huang & Chang, 2008), any proposed evaluation of books has largely been restricted to limited, partial applications using the traditional citation indexes. In 2009, the lack of more ambitious initiatives and alternative databases was challenged by a proposed set of indicators based on consulting the Online Public Access Catalogue (OPAC)[2] thanks, above all, to certain technological developments—such as the Z39.50 protocol—and, especially, the Online Computer Library Center's (OCLC) launch of WorldCat.org, in 2006. This open-access catalog unified in a single search engine millions of libraries enabling users to determine where any given title could be found (Nilges, 2006). The library count–based methodology was initially termed Library Catalog Analysis (LCA) or Library Holdings Analysis and was one of the first approaches to evaluation to challenge the use of citations; it was launched two years before the Altmetric manifesto (Priem et al., 2010) was published. Since then, the framework and methods enabling researchers to analyze the impact, diffusion and use of books, have broadened substantially.

Consequently, we currently have access to a broad-ranging set of indicators applicable to any document type, including all those generated on social media networks like Twitter, Wikipedia or newsfeeds (Torres-Salinas, Cabezas-Clavijo, & Jiménez-Contreras, 2013). Furthermore, other specific indicators have appeared and are unique to scientific books. These include the number of reviews recorded by the Book Review Index or the number and score available on the Goodreads or Amazon Reviews web platforms—the latter being related to popularity (Kousha & Thelwall, 2016). Similarly, in recent years, mentions in Syllabi, Mendeley bookmarks, or citations from audiovisual resources like YouTube (Kousha & Thelwall, 2015) have also been used. Moreover, to unify these metrics, platforms like Altmetric.com or PlumX Analytics have appeared—the latter pays special attention to the book since it integrates many earlier indicators, including library holdings. Therefore, sufficient sources of information about books and indicators of books are currently available.

So, bearing in mind the current surfeit of bibliometric resources, in the present chapter we seek to focus on the aforementioned Library Catalog Analysis (LCA) methodology—proposed by Torres-Salinas and Henk F. Moed—which could be considered one of the pioneering proposals in altmetrics, at least with reference to the evaluation of books. The objective of the present chapter is not just to pay tribute to Henk, it is also to offer a current perspective of library holdings–based indicators. The text has been organized in five parts. In the first, we present the origins of LCA and the first proposals, comparing their common characteristics and differences. The next section focuses on correlations with other indicators and discusses theories about their significance. We then continue with a critical analysis of the principal sources of information (WorldCat, PlumX Analytics, etc.). And finally, for the first time, we illustrate and apply the use of the WorldCat Identities tool[3] to the field of Informetrics in order to identify the principal authors and monographs.

---

[2] An OPAC is an online database that enables us to consult a library catalog.

## 2. The Origins of Library Catalog or Library Holdings Analysis

The seeds of collaboration between one of the present authors (D.T.-S.) and Henk F. Moed leading to the LCA proposal, date back to 2007 when, as a visiting researcher, Henk spent some time at the University of Granada (Spain)—at the invitation of the Grupo EC3 research group—in order to prepare the 11th ISSI 2007 conference in his role as program chair. During his stay, Henk and D.T.-S. discussed a range of topics relating to the latter's upcoming research visit to CWTS Leiden (The Netherlands) and decided to work on a new approach to the evaluation of the scientific book. Thus, the LCA proposal was born—to be further developed during D.T.-S.'s visit from October 2007 to February 2008. Initial results were presented at the 10th STI conference (Torres-Salinas & Moed, 2008) and the first draft paper was finally submitted to the *Journal of Informetrics* in August 2008; it was accepted in October 2008 and published online on December 30, 2008 (Torres-Salinas & Moed, 2009).

To better define that proposal, the present authors have recovered an e-mail message in which Henk explained LCA to Charles Erkelens, the then Editorial Director at Springer:

> *The aim of this project is to analyse the extent to which scientific/scholarly books published by a particular group of scientists/scholars are available in academic institutions all over the world, and included in the catalogs of the institutions' academic libraries. The base idea underlying this project is that one can obtain indications of the 'status', 'prestige' or 'quality' of scholars, especially in the social sciences and humanities, by analysing the academic libraries in which their books are available. We developed a simple analogy model between library catalog analysis and classical citation analysis, according to which the number of libraries in which a book is available is in a way comparable to the number of citations a document receives. But we realise that our data can in principle be used for other purposes as well. The interpretation of the library catalog data is of course a very complex issue.*
> *(Moed, Henk F. Personal communication - email, January 21, 2008).*

More precisely, in the 2009 study LCA was defined as "the application of bibliometric techniques to a set of library online catalogs". As a case study, Torres-Salinas & Moed selected the field of Economics and searched for books on the topic available in 42 university libraries in 7 different countries. They analyzed 121 147 titles included on 417 033 occasions in the sample libraries, making this one of the first bibliometric studies to use large-scale data about books. The authors proposed 4 indicators—the most noteworthy being the Number of Catalog Inclusions and the Catalog Inclusion Rate (Table 1)—and successfully extrapolated techniques like Multidimensional Scaling (MDS) and Coword Mapping. They then conducted a further two case studies which focused on the University of Navarra (Spain) and studied the major publishing houses in the field of Economics.

Fundamental to the development of their methodology were conversations with Adrianus J. M. Linmans who, in 2007-2008, was a member of the CWTS staff. University of Leiden librarian Linmans had also considered the use of catalogs as a tool to obtain quantitative data, especially in the field of the Humanities, and had conducted several applied studies the results of which had been presented internally at CWTS (Linmans, 2008). Part of these was subsequently published in *Scientometrics* in May 2010 (online in August 2009) (Linmans, 2010). The role of Henk F. Moed was also crucial to these contributions as Linmans himself explicitly acknowledged: "I am grateful to Henk Moed for his encouraging me to investigate library catalogues as a bibliometric source" (Linmans, 2010, p. 352).

---

[3] WorldCat Identities: https://worldcat.org/identities/

**Table 1. Main indicators for Library Catalog or Libcitations Analysis proposed in 2009**

| Indicator | Definition |
|---|---|
| **Proposed by Torres-Salinas & Moed** | |
| **CI** Catalog Inclusions | The total number of catalog inclusions for a given set of book title(s). This indicates the dissemination of a (given set of) book title(s) in university libraries. |
| **RCIR** Relative Catalog Inclusion Rate | This is defined as the ratio of CIR of the aggregate to be assessed and the CIR of the aggregate that serves as a benchmark in the assessment. A special case is the calculation of an RCIR in which the CIR of an institute under assessment is divided by the CIR calculated for the total database. A value above 1 indicates that an institution's CIR is above the world (or total database) average. |
| **DR** Diffusion Rate | The percentage of catalog inclusions of book titles produced by a given aggregate relative to the total number of possible catalog inclusions. The number of possible inclusions is equal to the product of the number of titles in the set and the number of catalogs included in the analysis. DR values range between 0 and 1. A value of 1 indicates that each title analyzed is present in all the library catalogs studied |
| **Proposed by White et al.** | |
| **Libcitations** Library Citations | For a particular book (i.e., edition of a title), this increases by 1 every time a different library reports acquiring that book in a national or international union catalog. Readers are invited to think of union catalogs in a new way: as "librarians' citation indexes". |
| **CNLS** Class Normalized Libcitation Score | To compute the CNLS, we obtained the number of books in each target item's LC [Library of Congress] class and the sum of libcitations of all those books. These data allowed us to compute the mean libcitations in each LC class as an expected value by which to divide the book's observed libcitation count. For example, if the mean libcitation count for an LC class is 20 and the book's libcitation count is 40, then CNLS=2, or twice the average for that LC class |
| **RC** Rank in Class | We also show each book's LC class and its rank in that class with respect to other titles. This measure resembles one already used in evaluative bibliometrics: the position of an author's or research unit's citation count in an overall distribution of citation counts. |

Linmans' contributions offer a different perspective to that of Torres-Salinas & Moed since he does not focus on a specific field but, rather, on the analysis of 292 lecturers ascribed to the Faculty of Humanities at the University of Leiden; hence the context and level of aggregation are totally different and more applied in nature. As well as library catalogs, Linmans made use of traditional indicators that enabled him to calculate the first correlations. Terminologically speaking, it should be noted that instead of Library Catalog Analysis he referred to his methodology as Library Holdings Analysis (in the present chapter these terms are treated as synonyms). With regard to the sample, Linmans analyzed 1135 books present in 59 386 book holdings in the United States, United Kingdom and The Netherlands, and employed the WorldCat collective catalog for his calculation. Linmans introduced interesting methodological variations in his calculations of indicators at author level: he distinguished types of book production by responsibility (editor or author), and differentiated between the language of publication of books.

However, these proposals were not unique since Howard D. White described a similar methodology in *JASIST*—published online in February 2009. This can only lead us to conclude that in 2008 both in Europe and in the United States researchers had been simultaneously working on the development of the same method in complete ignorance of each other. In fact, *JASIST* received White's paper on July 31, and the *Journal of Informetrics* received Torres-Salinas & Moed's submission on August 1. The phenomenon of simultaneous discovery—quite common in Science—was confirmed by White in his introduction: "*After this article had been submitted to* JASIST*, we learned that the same parallelism between citation counts and library holdings counts had been proposed in a conference paper by Torres-Salinas and Moed in 2008. The appearance of similar proposals in wholly independent projects suggests that this is an idea whose time has come*". (White et al., 2009, p. 1084).

**Table 2. Principal characteristics of the three studies of Library Holdings published simultaneously**

| Bibliographic Reference | Publication history | Denomination and definition | Level of analysis |
|---|---|---|---|
| Torres-Salinas, D., & Moed, H. F. (2009). Library Catalog Analysis as a tool in studies of social sciences and humanities: An exploratory study of published book titles in Economics. Journal of Informetrics, 3(1), 9-26. | *Journal of Informetrics*: <br> - Received: 1 August 2008 <br> - Accepted: 22 October 2008 <br> - Published online: 30 December 2008 | **Library Catalog Analysis** <br> The application of Bibliometric techniques to a set of online library catalogs. In this paper LCA is used to describe quantitatively a scientific–scholarly discipline and its actors, on the basis of an analysis of published book titles. | Analysis by discipline <br><br> Field analyzed: Economics |
| White, H. D. et. al. (2009). Libcitations: A measure for comparative assessment of book publications in the humanities and social sciences. *Journal of the American Society for Information Science and Technology*, *60*(6), 1083-1096. | *JASIST*: <br> - Received:30 July 2008 <br> - Accepted: 9 January 20098 <br> - Published online: 20 February 2009 | **Libcitation analysis** <br> The idea is that, when librarians commit scarce resources to acquiring and cataloging a book, they are in their own fashion citing it. The number of libraries holding a book at a given time constitutes its libcitation count. | Author level <br><br> Field analyzed: History, Philosophy, and Political Science, |
| Linmans, A. (2010). Why with bibliometrics the humanities does not need to be the weakest link: Indicators for research evaluation based on citations, library holdings, and productivity measures. *Scientometrics*, *83*(2), 337-354. | *Scientometrics* <br> - Received: 28 January 2009 <br> - Published online: 13 August 2009 | **Library holdings analysis** <br> A set of impact indicators, measuring the extent to which books by the same authors are represented in collections held by representative scientific libraries in different countries. | Faculty level <br><br> Field analyzed: Humanities and Social Sciences |

White's proposed methodology is theoretically and practically the same as that of LCA, or Library Holdings Analysis, although White did choose the elegant term "libcitations" to describe the number of library holdings in which a book is found. Undoubtedly, White's term for the new indicator seems better suited than Torres-Salinas & Moed's "Library Inclusions" and we cannot help but recognize it as being more descriptive and more appropriate. Methodologically, White's proposal is more akin to Linmans' approach since, firstly, faced with a macro- or discipline-oriented perspective, he too focuses at micro level on the production of 148 authors from different departments (Philosophy, History and Political Science) of various Australian universities (New South Wales, Sydney); secondly, also like Linmans, he chose WorldCat as his source of information. However, in relation to the indicator he does have more in common with Torres-Salinas & Moed in designing more complex indicators such as the Class Normalized Libcitation Score (CNLS) which facilitates a contextualization of the results and is much like these authors' proposed Relative Catalog Inclusion Rate (Table 2).

We conclude that the LCA was, at that time, a new methodology that offered a quantitative vision and an alternative narrative to the traditional bibliometric indicators. So, it was born as a surprising, simultaneous, triple proposal and drew essentially on the technological change of the period with the creation of new sources of information, in this case WorldCat. Over the last 10 years, limited but determined research interest has centered on the evaluation of the scientific book, now integrated into the universe of altmetrics where the methodology has been tested in different ways (Zuccala et al., 2015; Biagetti, 2018b). In the coming sections we will focus on some of these aspects, especially in relation to other indicators and the sources of information available when undertaking to analyze library catalogs.

## 3. Correlations and meaning

As is usually the case when a new indicator takes to the stage, studies that analyze its correlations with other metrics tend to abound. Library Holdings Counts, or inclusions, are no exception to the rule and in the light of the results we can confirm that they offer a different view to that of citation indicators. Although research into the relation between citations and libcitations has almost always used different methods, sources of information and disciplines, a pattern does appear: correlations, although occasionally significant, are usually low and of little relevance as, for example, in the work of Linmans (2010) and Zuccala and Guns (2013). Everything suggests that citations and libcitations do capture certain information in common but no cause-effect relation appears to exist in either direction. We are faced with an indicator that measures or depicts a type of impact or diffusion that is different to that of the citation.

Among the studies that confirm these facts (Table 3), the first is that by Linmans (2010) which tackles the correlations between library holdings and citations calculated from the Web of Science in the context of the University of Leiden's Faculty of Humanities. The correlation Linmans calculated from his data set was 0.29, rising to 0.49 for English language books. Zuccala & White (2015) reported on data for two disciplines and two time spans, distinguishing between centers that belonged to the Association of Research Libraries (ARL) and those that did not. Their analysis covered the period 1996-2011, the Scopus database, and 10 disciplines within the Humanities. For the two major disciplines, History and Literature, they obtained correlations of 0.24 and 0.20, respectively; when limiting their study to ARL centers, these rose to 0.26 and 0.24, respectively. In general, if we consider the 10 disciplines and two time spans, correlations rarely exceed 0.20, and only exceptionally reach 0.28.

**Table 3. Reported correlations between the number of inclusions in library holdings, or libcitations, and the number of citations**

| Author | Study type | Citation database | Correlation coefficient |
|---|---|---|---|
| Linmans (2010) | Published books in a Faculty of Humanities library | Web of Science | 0.29 All books<br>0.40 Books in English |
| Kousha & Thewall (2016) | 759 books in Social Sciences and 1262 in the Humanities | Book Citation Index | 0.145 Social Sciences<br>0.141 Humanities |
| Kousha & Thewall (2016) | 759 books in Social Sciences and 1262 in Humanities | Google Books | 0.234 Social Sciences<br>0.268 Humanities |
| Zuccala & White (2015) | 20 996 books in History and 7541 in Literature & Literary Theory cited in Scopus journals for 2007-2011 | Scopus | 0.24 History<br>0.20 Literature & Literary Theory |
| Zhang, Zhou, & Zhang (2018) | 2356 indexed in the Chinese Social Sciences Citation Index | Chinese Social Sciences Citation Index | 0.291 Ethnology<br>< 0.20 Other disciplines |

Kousha and Thewall (2015) used sources like the Book Citation Index and Google Books and reported higher correlations in the latter: 0.234 in Social Sciences and 0.268 in the Humanities. In other linguistic contexts correlations have not been particularly positive either. Zhang, Zhou and Zhang (2018) calculated the citation correlations of 2356 books included in the Chinese Social Sciences Citation Index with their respective inclusions in library holdings (OCLC) in 21 scientific disciplines. Among these, Ethnology attained the highest score with a Spearman correlation coefficient of 0.291; other areas were systematically below 0.2.

**Figure 1. Spearman correlation coefficients between several PlumX Analytics indicators and Library Holdings Counts indicators.**

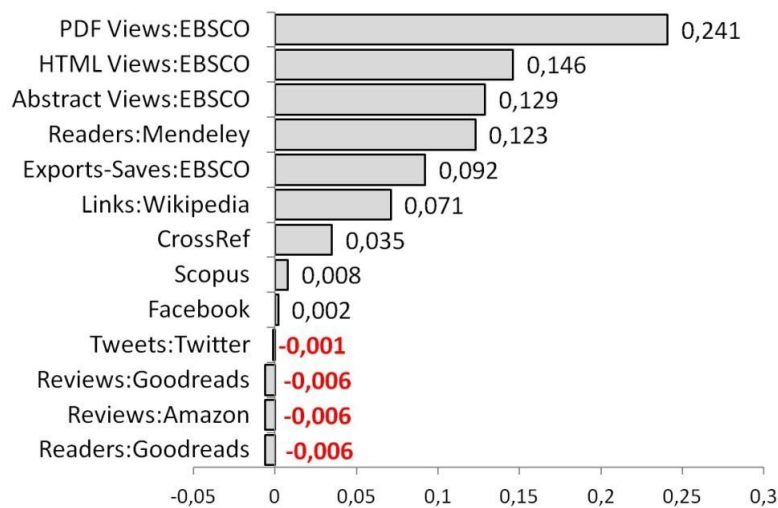

Note: this figure has been constructed using data from complementary materials published in the analysis of PlumX Analytics by Torres-Salinas, Gumpenberger & Gorraiz (2017).

Other studies have looked at the relation between Library Holdings and altmetrics. Kousha and Thewall (2016) analyzed the libcitation correlations of a set of books found in the *Books Citation Index* with various Amazon indicators (reviews, stars, sales rank, etc.). Both in Social Sciences (n=759) and the Humanities (n=1262), the best correlations with Amazon Reviews were 0.321 and 0.348, respectively. In Engineering (n=718) all correlations were substantially lower and 0.129 was the highest value for the reviews. Torres-Salinas, Gumpenberger and Gorraiz (2017) also reported a low correlation between library counts and other altmetrics included in PlumX Analytics (Figure 1). In this case, the sample was a collection of 268 061 books and the higher correlations were with three indicators related to visits to the EBSCO platform (PDF views=0.241, html=0.146, Abstract=0.129).

In the light of these results, we clearly find ourselves faced with an indicator that measures a dimension other than that of scientific impact. Kousha & Thewall (2016) point to libcitations' capacity to capture the educational value[4] of books and their cultural influence. Other authors offer a simplified view of the issue indicating that presence in libraries reflects the positive or negative reception of a book (Biagetti, 2018b) but that this is not a measure of scientific evaluation. In our view, when studying academic libraries, the Library Inclusions or Counts indicator expresses the potential value of a book to the scientific community (Torres-Salinas, 2009). This opinion coincides with Zuccala and White (2015) who affirm that while inclusion in libraries reflects "the potential readerships or users of a given book", it is not a reflection of social impact, even though books that have produced social change do attain high scores on this indicator (White et al. 2009; Zuccala & White 2015). Whatever the case may be, significance clearly does change in line with the sample of libraries used. If we only analyze indexations of books in public library collections, the significance of libcitations may more closely reflect cultural impact; if we analyze academic libraries, it may come closer to reflecting educational use and impact.

Another issue that could question the value and significance of these indicators is the fact that books found in a catalog do not always respond to a librarian's choice since many are donations, gifts, and so on. Biagetti (2018b) is highly critical on this issue, indicating that books are not always chosen consciously and describing how many purchases are predetermined by cooperative acquisition, purchasing plans, or individuals' requests. Biagetti (2018a), states that donations are one of the most critical issues and reports that 4.2% of books (n=2165) in Bibliography and Library Science and 8.2% (n=1097) in History of Political Institutions were donations. In the light of these results, we should not consider donations as a problem but, rather, treat them as a factor similar to self- or negative citation. Similar criticisms can be found in Hammarfelt (2016, p. 122) who reports that library purchases made "on automatic pilot" imply no merit on the part of the author. However, perhaps the factor that most distorts the value of libcitations—as a consequence of the selection process—is the purchase of e-books, since these are integrated into library collections *en masse* through packs and databases (Lewis & Kennedy, 2019).

Whatever the case may be, beyond the doubts raised by the acquisition process, we believe that Library Holdings do capture the potential use of a book. Specialized libraries acquire books for their users (students and researchers). This implies a value judgment that is usually the result of a Selection Plan or of Patron-Driven Acquisitions (PDA) (Tyler et al., 2019); i.e. these are acquisitions made as a function of the real use that they may have (Yuan, Van Ballegooie, & Robertson, 2018). When a book is found in many libraries that implies at least two things: 1) the book has been perceived to be of value either by the librarian or by a user (PDA or librarians' orders) and 2) the book, because it is held in many libraries, has a readership and a potential use greater than those of books that are not there.

---

[4] Table 8 illustrates our confirmation or perception of the educational value of books—a dimension not captured by citation since manuals or handbooks are rarely cited in scientific literature.

Furthermore, the type of use (cultural, scientific or educational in nature) is determined by the type of library we use as a source of information. Notwithstanding, we should say that Library Holding Inclusions is an indicator that should not be used in isolation to determine scientific impact, as does occur with other altmetrics (Haustein, 2016).

## 4. Sources of information

Firstly, we should consider that the LCA does not have a uniform methodology and each analysis entails the design of a specific methodology. As mentioned earlier, the most important factor is how the libraries are selected as this will determine the results and their interpretation. So, when it comes to dealing with a study of library holdings we have to consider three issues related with the sources of information:

a) Which libraries are we going to consult?
b) Which sources of information are we going to employ?
c) How are we going to gather the information?

In relation to libraries, most studies limit the search for books to a specific set within the OPACs. The selective use of particular catalogs as a function of certain characteristics (institutional prestige, linguistic context, etc.) is a question that arose in the initial studies. For example, Torres-Salinas & Moed (2009) selected only 42 university libraries considered prestigious in the field of Economics. However, if we were dealing with an analysis in a context that is non–English speaking (Spain, Italy, China, or wherever) or in disciplines with contents of a highly local nature (Archeology, History, etc.), we would be advised to select catalogs from the country itself or from the same geographical context. This is Biagetti's approach (2018a) in setting up a balanced mixed sample of catalogs from around the world but including 13 Italian libraries. Linmans (2010), however, selected libraries from just three countries: the USA, the UK and The Netherlands. Similarly, when using generic collective catalogs we should eliminate certain library types if we are seeking to analyze scientific impact. Hence, Zuccala and White (2015) distinguished libcitations as a function of the source library—in their case, whether or not this belonged to the ARL. These authors justified this by signaling that contributing to and helping Science is not a primary objective of public libraries.

With regard to primary sources of information, researchers have been using collective library catalogs for some years as these enable them to search a range of libraries from a single platform thanks to the Z39.50 protocol. Collective university and specialized library catalogs are frequent at national level; for example the *Jisc Library Hub Discover* in the UK[5] and Ireland, *SUDOC*[6] in France, *REBIUN*[7] in Spain, *INKA* in Germany[8] or *LIBIS* in Belgium[9]. These OPACS identify which libraries index a given book and are useful sources for small-scale national and local studies. However, the importance and use of collective national catalogs has been mitigated by the OCLC's WorldCat which is capable of agglutinating thousands of catalogs worldwide. Hence, it is hardly surprising that WorldCat has become the database *par excellence* and dominates library catalog studies (Linmans, 2010; Zuccala & White, 2015; Neville & Henry, 2014; Halevi, Nicolas, & Bar-Ilan, 2016).

---
[5] Jisc Library Hub Discover: https://discover.libraryhub.jisc.ac.uk/
[6] SUDOC - Catalogue du Système Universitaire de Documentation: http://www.sudoc.abes.fr/xslt/
[7] REBIUN - Red de Bibliotecas Universitarias Españolas: https://www.rebiun.org
[8] INKA - Inkunabelkatalog: http://www.inka.uni-tuebingen.de
[9] LIBISnet: http://libis.be/libis/libisnet

**Table 4. User location and number of academic libraries in WorldCat**

| Countries | Users Location[note1] | Number of academic libraries[note2] | Number of public libraries[note2] | Number of other types of library[note2] | Total number of libraries[note2] |
|---|---|---|---|---|---|
| United States | 44.8% | 2505 (43.16%) | 3532 (79.53%) | 3478 (70.26%) | 9515 (62.62%) |
| United Kingdom | 3.7% | 137 (2.36%) | 131 (2.95%) | 88 (1.78%) | 356 (2.34%) |
| Germany | 3.2% | 307 (5.29%) | 18 (0.41%) | 89 (1.8%) | 414 (2.72%) |
| France | 2.3% | 1113 (19.18%) | 9 (0.2%) | 34 (0.69%) | 1156 (7.61%) |
| Italy | 1.7% | 104 (1.79%) | 113 (2.54%) | 20 (0.4%) | 237 (1.56%) |
| Spain | 1.5% | 42 (0.72%) | 6 (0.14%) | 19 (0.38%) | 67 (0.44%) |

[note1] Geographical location of users, results from log (Wakeling et al., 2017)
[note2] Data from the Directory of OCLC members

One reasons for WorldCat's success is its size. The OCLC directory currently identifies 15 194 libraries, 5804 of which are academic[10]. However, despite WorldCat's obvious advantages, few studies have critically studied its use even though it has a clear English language bias (Wakeling et al., 2017). Table 4 shows that 44.8% of platform users and 43% of academic libraries are from the US—much higher figures than those of any European country. For example, only 1.5% and 0.7% of the libraries are Spanish, which undoubtedly forces Spanish researchers to use complementary sources when analyzing the diffusion of books in Spain—of 76 Spanish university libraries, only 42 are present in WorldCat. So, we would recommend that when conducting an LCA with WorldCat, researchers should consult the OCLC members' directory to verify the catalogs' territorial distribution.

The third question we should deal with is how we are going to consult the libraries and gather data. One approach is to consult the catalogs via a generic search engine but this implies performing manual searches or using a technique such as web scraping. Another option is the recovery of bibliographic data via a Z39.50 client, among which we would highlight Bookwhere Suite by WebClarity[11]. However, only Torres-Salinas & Moed (2008) have used this approach which, although it facilitates the creation of a large database with a high level of normalization, it also entails problems: the Z39.50 protocol needs to be established manually for each library; not all libraries share their connection data publicly; connection errors; and so on. Consequently, this approach is seldom used and studies of this type usually employ catalog APIs.

Many individual centers have APIs, for example Harvard Library[12], as do collective catalogs like, for example, COPAC[13]. WorldCat also has an API[14], which is useful for this type of study since searches can be launched using a range of parameters (ISBN, OCLC number, etc.)

---

[10] Information drawn from the Directory of OCLC Members: https://www.oclc.org/en/contacts/libraries.html. Note that some OCLC sources put the number of member libraries at 17 983: https://www.oclc.org/en/about.html
[11] Web site: http://www.webclarity.info/products/bookwhere/
[12] Harvard Library APIs & Datasets: https://library.harvard.edu/services-tools/harvard-library-apis-datasets
[13] COPAC API: https://www.programmableweb.com/api/copac
[14] WorldCat Search API: https://platform.worldcat.org/api-explorer/apis/wcapi

and data obtained on libraries and their locations. Results are returned by the API in MARC XML or Dublin Core formats for bibliographic records, and in XML or JSON for the library catalog URL and library locations (Table 5), with an upper limit of 50 000 consultations/day. In return, access is granted to a member institution that supervises and is responsible for data use[15], which makes it less than totally free and independent for researchers.

Table 5. Supported Operations for library locations in the WorldCat API

| Library Locations | Information offered by WorldCat | URL path |
|---|---|---|
| GetByOCLCNumber | Get Library Locations by OCLC Number and geographic information | /content/libraries/{OCLC_Number} |
| GetByISBN | Get library locations by ISBN near a geographic location | /content/libraries/isbn/{ISBN} |
| GetByISSN | Get library locations by ISSN near a geographic location | /content/libraries/issn/{ISSN} |
| GetByStandardNumber | Get library locations by Standard Number and geographic location | /content/libraries/sn/{Standard_Number} |

Those sources that have gained an upper hand in the world of altmetrics are aggregators capable of recovering a set of indicators for a set of document by simply entering any normalized number (DOI, Handle, etc.). Due to the special attention that it has paid to the altmetrics of books and monographs, the aggregator PLumX Analytics is outstanding (Torres-Salinas, Gumpenberger & Gorraiz, 2017). In this it contrasts with the relative inefficiency of Altmetric.com when evaluating books, as revealed in recent studies (Torres-Salinas, Gorraiz, & Robinson-García, 2018). PlumX, currently owned by Elsevier, includes Library Holdings Inclusions in WorldCat (Holdings: WorldCat) among its indicators and easily enables its calculation from the ISBN and searches on a vast scale. One of the first studies to use this source was Halevi, Nicolas and Bar-Ilan (2016) which used 71 443 eBook ISBN numbers from the Levy Library Ebrary collection to create a ranking of the books with most inclusions, 98.80% of which were indexed in WorldCat.

In consonance with earlier studies, Torres-Salinas, Gumpenberger & Gorraiz (2017a) also worked with PlumX and a set of 263 210 libros, of which 97.81% were included in at least one library, 18 indicators were studied and WorldCat in PlumX had the best coverage and the highest values (Table 6). Other studies have confirmed the value of PLumX—for example, Torres-Salinas, Robinson-García, & Gorraiz (2017b). This study is relevant since it analyzed a sample (n=2957) of books published by University of Granada researchers, finding that 48% of the metric hits for books corresponded to Library Holdings Inclusions and 79% of the books presented this metric. The study shows both the success of PlumX in including WorldCat–based indicators and the potency of the library holdings counts versus other altmetrics indicators.

---

[15] WorldCat Search API for OCLC Member Institution: Terms and Conditions:
https://www.oclc.org/content/dam/developer-network/PDFs/wcapi-terms-and-conditions-20121204.pdf

**Table 6. Coverage of indicators included in the PLumX aggregator and reported elsewhere**

| Indicators | Results from Halevi* | Results from Torres-Salinas** |
|---|---|---|
| WorldCat Holdings | 98.80% | 97.81% |
| Abstract views: EBSCO | 91.52% | 95.12% |
| Saves: EBSCO | 78.19% | 81.77% |
| PDF views: EBSCO | 65.64% | 68.28% |
| Goodreads captures | 69.23% | 53.70% |
| Mendeley captures | 43.11% | 24.86% |
| Goodreads reviews | 25.08% | 19.13% |
| Wikipedia links | 25.47% | 16.57% |
| Scopus or CrossRef Citations | 4.04% | 4.25% |

*Halevi: 2016. n=71 443. Levy Library Ebrary collection
**Torres-Salinas 2017. n=263 210. University of Vienna, the EBSCO e-book collection

In our final methodological considerations we would like to stress that when conducting an LCA the formal characteristics of books themselves must be considered, above all when using large-scale data collection methods: individual titles may appear with different ISBNs because they are in different formats or editions; translations into other languages may exist; or national cataloging procedures may differ (Biagetii, 2018a). All of these issues, already highlighted in the earlier studies (Linmans, 2010), cause difficulties when matching titles and ISBNs and, above all, affect collective catalogs that integrate highly heterogeneous information; such problems are reflected, for example, in differences in library counts between the WorldCat home page, the API, or WorldCat Identities.

## 5. An experiment with WorldCat Identities

One level of aggregation at which Library Holdings Counts has yet to be seen to be applicable is that of author. Similarly, despite its evident potential, it has yet to be used in a WorldCat Identities study. In the present section, we seek to remedy this by applying LCA at the level of author and discipline—specifically Informetrics—using WorldCat Identities. Thanks to a range of data mining and clustering techniques[16], WorldCat Identities unifies under a single normalized heading the "complete works" of any given author and calculates library diffusion data of that work both at author level and by individual publication; moreover, it integrates context-based data (genre, topics, name variants, co-authors, etc.). For example, Figure 2 shows the complete record of information for Henk F. Moed.

---

[16] WorldCat Identities has 30 million entries and groups together information from sources such as VIAF and FAST. More information https://www.oclc.org/research/themes/data-science/identities.html

**Figure 2. Basic information offered by WorldCat Identities for a given author**

## Moed, H. F.

### Overview

| | |
|---|---|
| Works: | 45 works in 165 publications in 5 languages and 2,385 library holdings |
| Genres: | Handbooks and manuals  Conference papers and proceedings |
| Roles: | Author, Editor, Other, Creator |
| Classifications: | PN171.F56, 001.42 |

So, taking WorldCat Identities as our source, we have consulted a representative sample of 22 renowned researchers in the field of Informetrics. For each of them, we collected grouped data on WorldCat Identities about Library Holdings and Publications/Works, and complemented this with their Google Scholar profile citation numbers. These results are in Table 7. The author with most Library Holdings Counts for their entire work is Blaise Cronin with a total of 6749, followed by Chaomei Chen and Leo Egghe, with 5867 and 3718, respectively. Henk Moed occupies fifth position with 2385.

The most frequently cited authors do not always occupy the higher places in the libcitation classification. For example, Loet Leydesdorff, Mike Thelwall or, most clearly, Ton van Raan (406 libraries; 13 857 citations). Evidently this list is not based on scientific articles but on monographs, conferences or works in which the researcher appears as the contributor of a chapter. Hence the Library Holdings ranking gives special relevance to authors or editors of handbooks or manuals. This classification reflects one aspect based on another type of activity and academic contributions related to the generation of teaching/educational contents (e.g. manuals and professional books) or the author's engagement in their field (e.g. editing conference proceedings). Clearly some authors both contribute to a field and undertake other activities or publish other materials beyond articles. For example, note the profiles of Chaomei Chen or Henk Moed himself. The classification, as we stated earlier, captures the value of an academic activity beyond the citation.

We wish to complete this profile of Informetrics with Library Holdings Counts through WorldCat Identities with a list of the outstanding books that have led us to construct Table 7 and enabled us to distinguish two contrasting phenomena. Firstly, we have a set of books of great scientific impact and diffusion in libraries. In this context, the outstanding title is Henk F. Moed's *Citation analysis in research evaluation*; it is the most cited (1711 citations in Google Scholar) and third ranked in the highest number of libraries, 1010. Other similar titles are Chen's *Mapping scientific frontiers: the quest for knowledge visualization* (808 libraries; 401 citations) or De Bellis's *Bibliometrics and citation analysis: from the Science Citation Index to cybermetrics* (509 libraries; 728 citations). These books enjoy universal scientific recognition and, moreover, are reference manuals or handbooks—a value that in itself captures the library counts indicator.

**Table 7. List of researchers in the field of Informetrics ordered by their WorldCat Identities Library Holdings number**

| Author Worldcat Entities | Number of Works & Publications Worldcat Entities[17] | Library Number Holdings | Google Scholar Citations |
|---|---|---|---|
| Cronin, Blaise | 144 works in 582 publications | 6749[18] | 11 122 |
| Chen, Chaomei | 42 works in 243 publications | 5867 | 15 968 |
| Egghe, L. (Leo) | 57 works in 186 publications | 3718 | -- |
| Garfield, Eugene | 150 works in 447 publications | 3386 | 30 105 |
| Moed, H. F. | 45 works in 165 Publications | 2385 | 13 026 |
| Sugimoto, Cassidy R. | 10 works in 85 publications | 2270 | 5894 |
| Braun, Tibor | 156 works in 389 publications | 2268 | 9430 |
| Wolfram, Dietmar | 15 works in 49 publications | 1769 | -- |
| Debackere, Koenraad | 105 works in 175 publications | 1628 | 9762 |
| Ingwersen, Peter | 33 works in 142 publications | 1608 | 11 316 |
| Rousseau, R. | 25 works in 121 publications | 1385 | 13 534 |
| Rowlands, Ian | 22 works in 92 publications | 1298 | 5188 |
| Leydesdorff, L. A. | 64 works in 189 publications | 1230 | 47 889 |
| Thelwall, Mike | 46 works in 113 publications | 1132 | 28 585 |
| Glänzel, Wolfgang | 53 works in 114 publications | 1115 | 18 238 |
| De Bellis, Nicola | 7 works in 25 publications | 762 | -- |
| Narin, Francis | 45 works in 96 publications | 426 | 15 324 |
| Raan, A. F. J. van | 32 works in 68 publications | 406 | 13 857 |
| Schubert, András | 21 works in 62 publications | 394 | 12 816 |
| Persson, Olle | 121 works in 174 publications | 257 | 5205 |
| Bornmann, Lutz | 14 works in 28 publications | 215 | 14 945 |
| Nederhof, A. J. | 38 works in 59 publications | 199 | -- |

A second but more controversial group contains books that are present in many libraries but which have few citations: e.g. Holmberg's *Altmetrics for information professionals: past, present and future* (745 libraries; 41 citations); or Ding et al. *Measuring scholarly impact: methods and practice* (428 libraries; 2 citations). They clearly have a professional, practical profile, are not oriented towards a scientific readership and, thanks to the library counts, can now be analyzed from a different standpoint. Independently of their positions in Table 8, specialists in Informetrics will not be surprised, nor could anyone deny, that all these titles are key references in the field, whether from a scientific or a teaching/professional point of view.

Finally, and given that the present book is a tribute to Henk F. Moed, we must underline his contribution to the field since three of his books appear among the most popular works in Informetrics in library collections. Together with the abovementioned *Citation analysis in*

---

[17] The publication indicator should be read with caution since any given work may exist in different editions. Different editions can be distinguished as a function of the number of editions, format types (print or ebook) or, even, the same work catalogued differently in different libraries; WorldCat takes all these to be different "editions".

[18] To calculate this indicator we add in inclusions in tribute *Festschrift* publications. For example, in the case of Cronin we include 921 inclusions received by the book *Theories of informetrics and scholarly communication: a* Festschrift *in honor of Blaise Cronin*

*research evaluation* we find the *Handbook of quantitative science and technology research* (832 libraries), which he edited, and his latest monograph *Applied evaluative informetric* (298 libraries). If Henk can be said to stand out for anything it has been his contributions to the development of our discipline (Crown Indicator, SNIP, etc.) and, especially, in the most recent stage of his academic career, for making bibliometry more accessible to professional circles through manuals that we can already consider classic reference books.

**Table 8. Books about Informetrics with greater diffusion in WorldCat member libraries**

| Bibliographic Reference[19] | WorldCat member libraries | Google Scholar Citations |
|---|---|---|
| **Egghe, Leo.** *Power laws in the information production process: Lotkaian informetrics.* Amsterdam: Elsevier/Academic Press, 2005. | 1255 | 322 |
| **Wolfram, Dietmar.** *Applied informetrics for information retrieval research.* Westport, Conn. : Libraries Unlimited, 2003. | 1166 | 76 |
| **Moed, Henk.** *Citation analysis in research evaluation.* Dordrecht: Springer, 2005. | 1010 | 1711 |
| **Sugimoto, Cassidy R** (editor). *Theories of informetrics and scholarly communication: a Festschrift in honor of Blaise Cronin.* Berlin: De Gruyter, 2016 | 921 | 11 |
| **Moed, Henk; Glänzel, Wolfgang; Schmoch, Ulrich** (editors). *Handbook of quantitative science and technology research: the use of publication and patent statistics in studies of S & T systems.* Dordrecht Springer, 2011 | 832 | 328 |
| **Chaomei, Chen.** *Mapping scientific frontiers: the quest for knowledge visualization.* London: Springer, 2013 | 808 | 441 |
| **Holmberg, Kim.** *Altmetrics for information professionals: past, present and future.* Waltham: Chandos, 2016 | 745 | 41 |
| **Garfield, Eugene**. *Citation indexing - its theory and application in science, technology, and humanities.* Philadelphia: ISI Press, 1983 | 686 | 2924 |
| **Chaomei, Chen**. *CiteSpace: a practical guide for mapping scientific literature.* New York: Nova Science Publishers, Inc., 2016 | 571 | 36 |
| **Cronin, Blaise.** *The hand of science: academic writing and its rewards.* Lanham, Md: Scarecrow Press, 2005. | 515 | 267 |
| **De Bellis, Nicola**. *Bibliometrics and citation analysis: from the Science Citation Index to cybermetrics.* Lanham, Md: Scarecrow Press, 2009. | 509 | 728 |
| **Cronin, Blaise.** *Beyond bibliometrics: harnessing multidimensional indicators of scholarly impact.* Cambridge: The MIT Press, 2014 | 502 | 121 |
| **Ding, Ying; Wolfram, Dietmar; Rousseau, Ronald** (editors). *Measuring scholarly impact: methods and practice.* Cham: Springer, 2014 | 428 | 2 |
| **Leydesdorf, Loet; Besselaar, Peter Van Den** (editors). *Evolutionary economics and chaos theory: new directions in technology studies.* New York: St. Martin's Press, 1994. | 331 | 199 |
| **Thewall, Mike**. *Introduction to webometrics: quantitative web research for the social sciences.* San Rafael, Calif: Morgan & Claypool Publishers, 2009. | 328 | 375 |
| **Leydesdorf, Loet.** *Universities and the global knowledge economy: a triple helix of university-industry-government relations.* Pinter Pub Ltd 2005. | 306 | 67 |
| **Moed, Henk.** *Applied evaluative informetric.* Springer, 2017 | 298 | 43 |
| **Anne-Wil Harzing**. *The publish or perish book: your guide to effective and responsible citation analysis.* Melbourne, Australia: Tarma Software Research Pty Ltd, 2013. | 306 | 316 |
| **Vaan Raan, Anthony** (editor). *Handbook of quantitative studies of science and technology.* Amsterdam: Elsevier Science, 2014. | 232 | 273 |

## 6. Final remarks

In the present chapter we have sought to discuss several aspects of Library Catalog Analysis in our tribute to Henk F. Moed, who contributed intellectually to its conceptualization in one of the groundbreaking articles. In this sense, Moed supported and perceived the use of these new metrics at a time when indicators applied to the book were sorely lacking. Currently, library holdings—thanks above all to WorldCat and PlumX—constitute part of the altmetrics toolkit and, as they have shown in various studies, represent one of the indicators that offers the best coverage when compared with others such as reviews in Goodreads or Amazon. Thus, Library Holdings Inclusions constitute an ideal complement to combine with citations. As we have seen in the context of Informetrics, they reflect a professional or educational use and are especially valuable in analyzing those monographs that are oriented towards a non-scientific readership.

---

[19] To construct this list we have considered the works with the highest library counts of the author included in Table 8. Similarly, we have conducted searches using keywords such as informetrics, bibliometrics, altmetrics, and citation analysis, which have enabled us to identify works such as those of Kim Holmberg, Anne-Wil Harzing or Nicola De Bellis.

## 7. References

Archambault, É., Vignola-Gagné, É., Côté, G., Larivière, V., & Gingrasb, Y. (2006). Benchmarking scientific output in the social sciences and humanities: The limits of existing databases. *Scientometrics*, *68*(3), 329–342. https://doi.org/10.1007/s11192-006-0115-z

Biagetti, M. T., Iacono, A., & Trombone, A. (2018a). Is the Diffusion of Books in Library Holdings a Reliable Indicator in Research Assessment? BT  - The Evaluation of Research in Social Sciences and Humanities: Lessons from the Italian Experience. In A. Bonaccorsi (Ed.) (pp. 321–343). Cham: Springer International Publishing. https://doi.org/10.1007/978-3-319-68554-0_14

Biagetti, M. T., Iacono, A., & Trombone, A. (2018b). Testing library catalog analysis as a bibliometric indicator for research evaluation in Social Sciences and Humanities. In *Challenges and Opportunities for Knowledge Organization in the Digital Age: Proceedings of the Fifteenth International ISKO Conference 9-11 July 2018 Porto, Portugal* (1st ed., pp. 892–899). Baden-Baden: Ergon-Verlag. https://doi.org/10.5771/9783956504211-892

Hammarfelt, B. (2016). Beyond Coverage: Toward a Bibliometrics for the Humanities BT  - Research Assessment in the Humanities: Towards Criteria and Procedures. In M. Ochsner, S. E. Hug, & H.-D. Daniel (Eds.) (pp. 115-131). Cham: Springer International Publishing. https://doi.org/10.1007/978-3-319-29016-4_10

Haustein, S. (2016). Grand challenges in altmetrics: heterogeneity, data quality and dependencies. *Scientometrics*, *108*(1), 413–423. https://doi.org/10.1007/s11192-016-1910-9

Halevi, G., Nicolas, B., & Bar-Ilan, J. (2016). The Complexity of Measuring the Impact of Books. *Publishing Research Quarterly*, *32*(3), 187–200. https://doi.org/10.1007/s12109-016-9464-5

Hicks, D. (1999). The difficulty of achieving full coverage of international social science literature and the bibliometric consequences. *Scientometrics*, *44*(2), 193–215. https://doi.org/10.1007/BF02457380

Huang, M., & Chang, Y. (2008). Characteristics of research output in social sciences and humanities: From a research evaluation perspective. *Journal of the American Society for Information Science and Technology*, *59*(11), 1819–1828. https://doi.org/10.1002/asi.20885

Kousha, K., & Thelwall, M. (2015). Web indicators for research evaluation: Part 3: books and non standard outputs. *El Profesional de La Información*, *24*(6), 724–736. https://doi.org/10.3145/epi.2015.nov.04

Kousha, K., & Thelwall, M. (2016). Can Amazon.com reviews help to assess the wider impacts of books? *Journal of the Association for Information Science and Technology*, *67*(3), 566–581. https://doi.org/10.1002/asi.23404

Lewis, R. M., & Kennedy, M. R. (2019). The Big Picture: A Holistic View of E-book Acquisitions. *Library Resources & Technical Services*, *63*(2), 160. https://doi.org/10.5860/lrts.63n2.160

Linmans, A. J. M. (2008). *Een exploratieve studie van de onderzoeksprestaties van de Faculteit Letteren aan de Universiteit Leiden* (in Dutch). Internal CWTS Report

Linmans, A. J. M. (2010). Why with bibliometrics the Humanities does not need to be the weakest link - Indicators for research evaluation based on citations, library holdings, and productivity measures. *Scientometrics*, *83*(2), 337–354. https://doi.org/10.1007/s11192-009-0088-9

Nederhof, A. J. (2006). Bibliometric monitoring of research performance in the social sciences and the humanities: A review. *Scientometrics*, *66*(1), 81–100. https://doi.org/10.1007/s11192-006-0007-2

Neville, T. M., & Henry, D. B. (2014). Evaluating Scholarly Book Publishers—A Case Study in the Field of Journalism. *The Journal of Academic Librarianship*, *40*(3), 379–387. https://doi.org/https://doi.org/10.1016/j.acalib.2014.05.005

Nilges, C. (2006). The Online Computer Library Center's Open WorldCat Program. Library Trends, 54(3), 430–447. https://doi.org/10.1353/lib.2006.0027

Priem, J., Taraborelli, D., Groth, P., & Neylon, C. (2010). Altmetrics: A manifesto. Retrieved from http://altmetrics.org/manifesto/

Torres-Salinas, D., Cabezas-Clavijo, Á., & Jiménez-Contreras, E. (2013). Altmetrics New Indicators for Scientific Communication in Web 2.0. *Comunicar*, *21*(41), 53–60. https://doi.org/10.3916/C41-2013-05

Torres-Salinas, D., Gorraiz, J., & Robinson-Garcia, N. (2018). The insoluble problems of books: what does Altmetric.com have to offer? *Aslib Journal of Information Management*, *70*(6), 691–707. https://doi.org/10.1108/AJIM-06-2018-0152

Torres-Salinas, D., Gumpenberger, C., & Gorraiz, J. (2017a). PlumX As a Potential Tool to Assess the Macroscopic Multidimensional Impact of Books. *Frontiers in Research Metrics and Analytics*, *2*(July), 1–11. https://doi.org/10.3389/frma.2017.00005

Torres-Salinas, D., & Moed, H. (2008). Library catalog analysis is a useful tool in studies of social sciences and humanities. In *A New Challenge for the Combination of Quantitative and Qualitative Approaches. 10th International Conference on Science and Technology Indicators*. Viena.

Torres-Salinas, D., & Moed, H. F. (2009). Library Catalog Analysis as a tool in studies of social sciences and humanities: An exploratory study of published book titles in Economics. *Journal of Informetrics*, *3*(1), 9–26. https://doi.org/10.1016/j.joi.2008.10.002

Torres-Salinas, D., Robinson-Garcia, N., & Gorraiz, J. (2017). Filling the citation gap: measuring the multidimensional impact of the academic book at institutional level17

with PlumX. *Scientometrics*, *113*(3), 1371–1384. https://doi.org/10.1007/s11192-017-2539-z

Tyler, D. C., Hitt, B. D., Nterful, F. A., & Mettling, M. R. (2019). The Scholarly Impact of Books Acquired via Approval Plan Selection, Librarian Orders, and Patron-Driven Acquisitions as Measured by Citation Counts. *College & Research Libraries; Vol 80, No 4 (2019): May*. https://doi.org/10.5860/crl.80.4.525

Wakeling, S., Clough, P., Silipigni Connaway, L., Sen, B., & Tomás, D. (2017). Users and uses of a global union catalog: A mixed-methods study of WorldCat.org. *Journal of the Association for Information Science and Technology*, *68*(9), 2166–2181. https://doi.org/10.1002/asi.23708

White, H. D., Boell, S. K., Yu, H., Davis, M., Wilson, C. S., & Cole, F. T. H. (2009). Libcitations: A measure for comparative assessment of book publications in the humanities and social sciences. *Journal of the American Society for Information Science and Technology*, *60*(6), 1083–1096. https://doi.org/10.1002/asi.21045

White, H. D., & Zuccala, A. (2018). Libcitations, worldcat, cultural impact, and fame. *Journal of the Association for Information Science and Technology*, *69*(12), 1502–1512. https://doi.org/10.1002/asi.24064

Yuan, W., Van Ballegooie, M., & Robertson, J. L. (2018). Ebooks Versus Print Books: Format Preferences in an Academic Library. *Collection Management*, *43*(1), 28–48. https://doi.org/10.1080/01462679.2017.1365264

Zhang, H., Zhou, Q., & Zhang, C. (2018). Multi-discipline correlation analysis between citations and detailed features of library holdings. *Proceedings of the Association for Information Science and Technology*, *55*(1), 946–947. https://doi.org/10.1002/pra2.2018.14505501188

Zuccala, A., & Guns, R. (2013). Comparing book citations in humanities journals to library holdings: Scholarly use versus "perceived cultural benefit" (RIP). In *14th International Society of Scientometrics and Informetrics Conference, ISSI 2013*(Vol. 1, pp. 353–360). Amsterdam.

Zuccala, A., Guns, R., Cornacchia, R., & Bod, R. (2015). Can we rank scholarly book publishers? A bibliometric experiment with the field of history. *Journal of the Association for Information Science and Technology*, *66*(7), 1333–1347. https://doi.org/10.1002/asi.23267

Zuccala, A., & White, H. D. (2015). Correlating libcitations and citations in the humanities with WorldCat and scopus data. In S. A.A., S. A.A.A., S. C., A. U., & T. Y. (Eds.), *15th International Society of Scientometrics and Informetrics Conference, ISSI 2015*(pp. 305–316). Royal School of Library and Information Science, University of Copenhagen, Birketinget 6, Copenhagen S, DK-2300, Denmark: Bogazici Universitesi.

## 1. Introducción

La evaluación de la actividad científica en el ámbito de las Ciencias Humanas y Sociales y, especialmente del libro académico, ha sido una de las asignaturas pendientes de la bibliometría cuyo contexto evaluativo, hasta hace bien poco, ha estado monopolizado por indicadores de citación y por las bases de datos de Thomson Reuters (Nederhof, 2006). De esta manera, pese a que casi todos los estudios manifestaban la importancia de los libros en la comunicación científica (Archambault et al., 2006; Hicks, 1999; Huang & Chang, 2008), la mayoría de las propuestas en torno a la evaluación del libro eran aplicaciones parciales de poco alcance basadas en los índices de citas tradicionales. La falta de iniciativas globales y de base de datos alternativas se vio enriquecida cuando en 2009 se propusieron una serie de indicadores basados en la consulta de *Online Public Access Catalogue* (OPAC)[20] gracias, sobre todo, a diversos desarrollos tecnológicos como el protocolo Z39.50 pero, especialmente, al lanzamiento en 2006 de WoldCat.org por parte de la Online Computer Library Center (OCLC). Este catálogo de libre acceso unificó en un buscador millones de bibliotecas de todo el mundo indicando en cuales de ellas podíamos localizar un determinada título (Nilges, 2006). La metodología basada en el conteo de bibliotecas se dio a conocer como Library Catalog Analysis (LCA) o Library Holdings Analysis y es una de las primeras alternativas evaluativas frente a las citas, lanzada dos años antes de que se promulgase el manifiesto altmetric (Priem et al., 2010). Desde entonces se ha ampliado sustancialmente el marco y los métodos a partir de los cuales se puede analizar el impacto, difusión y uso de los libros.

En consecuencia, en la actualidad contamos con un amplio conjunto de indicadores aplicables a cualquier tipología documental como todos aquellos que se generan desde medios sociales como Twitter, Wikipedia o noticias (Torres-Salinas, Cabezas-Clavijo, & Jiménez-Contreras, 2013). Pero más allá de éstas también han aparecido otras específicas y exclusivas de los libros científicos, entre las cuales se encuentran el número de críticas recogidas en el Book Review Index o el número y puntuación disponible en las plataformas web Goodreads y Amazon Reviews, encontrándose en estas últimas una relación con su popularidad (Kousha & Thelwall, 2016). De la misma manera, también se han venido empleando en estos años las menciones realizadas en Syllabi, Mendeley bookmarks o las citas procedentes de recursos audiovisuales como YouTube (Kousha & Thelwall, 2015). Para unificar todas las métricas han surgido además plataformas como *Altmetric.com* o *PlumX Analytics*, siendo ésta última la que ha prestado una especial atención al libro ya que integra muchos de los indicadores anteriores, incluidos los library holdings. Por tanto, en la actualidad los libros cuentan con suficientes fuentes de información e indicadores.

Pues bien, teniendo en cuenta el presente contexto de superávit bibliométrico en este capítulo nos queremos centrar en la mencionada metodología del Library Catalog Analysis (LCA) que propusimos junto a Henk F. Moed y que podríamos considerar una de las propuestas pioneras de las altmétricas, al menos en lo que a la evaluación del libro se refiere. El objetivo del capítulo no es solamente homenajear a Henk sino también ofrecer una perspectiva actual de los indicadores basados en los library holdings. El texto se ha organizado en cinco partes. En la primera se presenta el origen y las primeras propuestas que se realizaron comparando sus características comunes y sus diferencias. El siguiente apartado se detiene en la correlación con otros indicadores y se exponen diferentes teorías sobre su significado. Se continúa este punto con un análisis crítico de las principales fuentes de información (WorldCat, *PlumX Analytics*, etc…). Finalmente, se ilustra por primera vez y aplica al campo de la informetría la utilidad de la herramienta WorldCat Identities[21] para identificar sus principales autores y monografías.

---

[20] Un OPAC es una base de datos on-line que permite consultar el catálogo de una biblioteca.

## 2. Orígenes del Library Catalog o Library Holding Analysis

El germen de la colaboración con Henk F. Moed para realizar la propuesta de Library Catalog Analysis (LCA) podemos fecharla en el año 2007, cuando Henk realiza una estancia en la Universidad de Granada, al amparo del Grupo EC3, para preparar el *11th ISSI* 2007 como programme chairman. Durante ese período tuvimos oportunidad de discutir diversos temas para mi futura estancia como investigador en el CWTS y decidimos trabajar en la evaluación del libro científico desde una nueva perspectiva. Por tanto, la propuesta de Library Catalog Analysis nace y se desarrolla como resultado de una estancia de investigación durante los meses de Octubre de 2007 a Febrero de 2008. Los primeros resultados fueron presentados en la *10th STI* (Torres-Salinas & Moed, 2008) y, finalmente, el primer borrador del paper fue enviado en agosto de 2008 al Journal of Informetrics, se aceptó en octubre de 2008 y fue publicado online el 30 de diciembre de 2008 (Torres-Salinas & Moed, 2009).

Para definir mejor nuestra propuesta he rescatado un e-mail en el cual Moed explicaba en qué consistía a Charles Erkelens, en aquel director editorial en Springer:

> *The aim of this project is to analyse the extent to which scientific/scholarly books published by a particular group of scientists/scholars are available in academic institutions all over the world, and included in the catalogs of the institutions' academic libraries. The base idea underlying this project is that one can obtain indications of the 'status', 'prestige' or 'quality' of scholars, especially in the social sciences and humanities, by analysing the academic libraries in which their books are available. We developed a simple analogy model between library catalog analysis and classical citation analysis, according to which the number of libraries in which a book is available is in a way comparable to the number of citations a documents receives. But we realise that our data can in principle be used for other purposes as well. The interpretation of the library catalog data is of course a very complex issue.*
> *(Moed, HF. Personal communication - email, January 21, 2008).*

Más concretamente, en nuestro trabajo definimos LCA como "*the application of bibliometric techniques to a set of library online catalogs*". Como estudio de caso se seleccionó la economía y se buscaron los libros publicados sobre el tema en 42 bibliotecas universitarias de 7 países diferentes. En total se analizaron 121 147 títulos que habían sidos incluidos 417 033 ocasiones en las diferentes bibliotecas, por tanto, es uno de los primeros trabajos bibliométricos que utiliza datos sobre libros a gran escala. A nivel de indicadores se propusieron cuatro medidas entre las que destacamos especialmente el *Number of Catalog inclusions* y la *Catalog inclusion Rate* (Tabla 1) y extrapolamos con éxito diferentes técnicas como el Multidimensional Scaling (MDS) o el Coword Mapping. Se realizaron con ello dos estudios de caso aplicados a la Universidad de Navarra y se ofreció un estudio de las principales editoriales de economía.

---
[21] WorldCat Identities: https://worldcat.org/identities/

Para el desarrollo de toda la metodología fueron fundamentales las conversaciones que mantuvimos con Adrianus J. M. Linmans que en el momento de la estancia formaba parte del staff del CWTS. Linmans, bibliotecario de la Universidad de Leiden, también había vislumbrado las posibilidades de utilizar los catálogos como una herramienta que pudiera ofrecer indicios cuantitativos, especialmente en el área de las Humanidades, y realizó diferentes estudios aplicados cuyos resultados fueron presentados internamente en el CWTS (Linmans, 2008). Parte de los mismos fueron publicados con posterioridad también en la revista Scientometrics en mayo de 2010 (online en agosto 2009) (Linmans, 2010). En las contribuciones mencionadas la figura de Henk F. Moed también fue crucial tal y como demuestra la nota de agradecimiento de Linmans en la que se expresa en los siguientes términos "*I am grateful to Henk Moed for his encouraging me to investigate library catalogues as a bibliometric source*" (Linmans, 2010, p. 352).

**Table 1. Main indicators proposed for Library Catalog or Libcitations Analysis in 2009**

| Indicator | Definition |
|---|---|
| **Proposed by Torres Salinas and Moed** | |
| **CI** Catalog inclusions | The total number of catalog inclusions for a given set of book title(s). It indicates the dissemination of a (given set of) book title(s) among university libraries. |
| **RCIR** Relative catalog inclusion rate | It is defined as the ratio of CIR of the aggregate to be assessed and CIR of the aggregate that serves as a benchmark in the assessment. A special case is the calculation of a RCIR in which CIR of an institute under assessment is divided by the CIR calculated for the total database. A value above one indicates that an institution's Catalog Inclusion Rate is above world (or total database) average. |
| **DR** Diffusion rate | The percentage of catalog inclusions of book titles produced by a given aggregate relative to the total number of possible catalog inclusions. The number of possible inclusions is equal to the product of the number of titles in the set and the number of catalogs included into the analysis. DR values range between 0 and 1. A value of 1 indicates that each title analysed is present in all library catalogs taken into account |
| **Proposed by White et al.** | |
| **Libcitations** Library Citations | For a particular book (i.e., edition of a title), it increases by 1 every time a different library reports acquiring that book in a national or an international union catalog. Readers are invited to think of union catalogs in a new way: as "librarians' citation indexes." |
| **CNLS** Class Normalized Libcitation Core | To compute the CNLS measure, we also obtained the number of books in each target item's LC [Library of Congress] class and the sum of libcitations to all of those books. These data allowed us to compute the mean libcitations in each LC class as an expected value by which to divide the book's observed libcitation count. For example, if the mean libcitation count for a class is 20 and the book's libcitation count is 40, then CNLS=2, or twice the average for that LC class |
| **RC** Rank in Class | We also show each book's LC class and its rank in that class with respect to other titles. This measure resembles one already used in evaluative bibliometrics: the position of an author's or research unit's citation count in an overall distribution of citation counts. |

Las contribuciones de Linmans ofrecen una perspectiva diferente a la que nosotros planteamos ya que no se centra en un campo concreto sino en el análisis de 292 profesores adscritos a la Facultad de Humanidades de Universidad de Leiden, por tanto el contexto y el nivel de agregación es totalmente diferente y con un carácter más aplicado. Además de los catálogos de bibliotecas, Linmans hace uso de indicadores tradicionales que le permitieron realizar las primeras correlaciones. Desde el punto de vista terminológico hay que hacer notar que en lugar de Library Catalog Analysis se refiere a su metodología como Library Holdings Analysis (en este capítulo usamos ambos términos como sinónimos). En relación a la muestra, Linmans analizó un total 1135 libros que estaban presentes en 59 386 book holdings de Estados Unidos, Reino Unido y Países Bajos empleando para su cálculo el catálogo colectivo WorldCat. Linmans introdujo interesantes matices metodológicos a la hora de calcular los indicadores a nivel de autor como distinguir la producción de libros según la responsabilidad (editor o autor) o bien matizar los cálculos según el idioma de publicación de los diferentes libros.

Pero las propuestas mencionadas no fueron las únicas ya que Howard D. White propuso una metodología similar publicada online en la revista JASIST en febrero de 2009. Todo hace suponer que durante el año 2008 en Europa y Estados Unidos habíamos estado trabajando simultáneamente en el desarrollo del mismo método sin tener conocimiento el uno del otro. De hecho, JASIST recibió el paper de White el 31 de julio y el Journal of Informetrics nuestro artículo el 1 de agosto. Este fenómeno de descubrimiento simultáneo, bastante habitual en la ciencia, también lo confirma White en la introducción de su artículo "*After this article had been submitted to JASIST, we learned that the same parallelism between citation counts and library holdings counts had been proposed in a conference paper by Torres-Salinas and Moed in 2008. The appearance of similar proposals in wholly independent projects suggests that this is an idea whose time has come*". (White et al., 2009, p. 1084).

**Tabla 2. Características principales de los tres estudios sobre Library Holdings publicados simultáneamente**

| Bibliographic Reference | Publication history | Denomination and definition | Nivel de análisis |
|---|---|---|---|
| Torres-Salinas, D., & Moed, H. F. (2009). Library Catalog Analysis as a tool in studies of social sciences and humanities: An exploratory study of published book titles in Economics. Journal of Informetrics, 3(1), 9-26. | Journal of informetric:<br>- Received: 1 August 2008<br>- Accepted: 22 October 2008<br>- Published online: 30 December 2008 | **Library Catalog Analysis**<br>The application of Bibliometric techniques to a set of library online catalogs. In this paper LCA is used to describe quantitatively a scientific-scholarly discipline and its actors, on the basis of an analysis of published book titles. | Análisis disciplinar<br><br>Campo analizado: Economía |
| White, H. D. et. al. (2009). Libcitations: A measure for comparative assessment of book publications in the humanities and social sciences. *Journal of the American Society for Information Science and Technology*, *60*(6), 1083-1096. | JASIST:<br>- Received:30 July 2008<br>- Accepted: 9 January 20098<br>- Published online: 20 February 2009 | **Libcitation analysis**<br>The idea is that, when librarians commit scarce resources to acquiring and cataloging a book, they are in their own fashion citing it. The number of libraries holding a book at a given time constitutes its libcitation count. | Author level<br><br>Campo analizado: history, philosophy, and political science, |
| Linmans, A. (2010). Why with bibliometrics the humanities does not need to be the weakest link: Indicators for research evaluation based on citations, library holdings, and productivity measures. *Scientometrics*, *83*(2), 337-354. | Scientometrics<br>- Received: 28 January 2009<br>- Published online: 13 August 2009 | **Library holdings analysis**<br>Group of impact indicators, measuring the extent to which books of the same authors are represented in the collections of representative scientific libraries in different countries. | Faculty level<br><br>Campo analizado: Humanities and Social Sciences |

La metodología propuesta por White, en términos teóricos y prácticos, es la misma que la de Library Catalog Analysis or Library Holdings Analysis, sin embargo el propio White elige con mayor gracia el término de Libcitations para describir el número de library holdings en las que se encuentra un libro. Sin duda, esta denominación del nuevo indicador frente a la denominación de Library Inclusion empleada por Torres-Salinas y Moed parece haber tenido más fortuna y, es justo reconocer, que es más descriptiva y adecuada. A nivel metodológico la propuesta de White presenta más similaridad con la de Linmans ya que, en primer lugar, frente a una perspectiva macro o disciplinar, también su estudio se centra en un nivel micro analizando la producción de 148 autores de distintos departamentos (Filosofía, Historia y Ciencias Políticas) de distintas universidades australianas (New South Wales, Sidney); en segundo lugar, al igual que Linmans, también escoge WorldCat como fuente de información. Por otro lado, desde el punto de vista de los indicadores sí presenta más puntos en común con nuestro trabajo al diseñar indicadores más complejos como el *Class Normalized Libcitation Score* (CNLS) que permite contextualizar los resultados y que es muy similar a nuestra propuesta de *Relative Catalog Inclusion Rate* (Tabla 2).

Podemos concluir que el LCA fue, en su momento, una nueva metodología que ofrecía una visión cuantitativa y un relato alternativo a los tradicionales indicadores bibliométricos. Nace asimismo como una sorprendente triple propuesta simultánea y se nutre esencialmente del cambio tecnológico del momento con el nacimiento de nuevas fuentes de información, en este caso WorldCat. En estos diez años se ha creado un pequeño frente de investigación en torno a la evaluación del libro científico, ahora integrado dentro del universo de las altmetrics, que ha testado esta metodología desde diferentes puntos de vista (Zuccala et al., 2015; Biagetti, 2018b). En los siguientes apartados nos centraremos en algunos de estos aspectos, especialmente en la relación con otros indicadores y en las fuentes de información disponibles para la realización de los Análisis de Catálogos de Bibliotecas.

## 3. Correlaciones y significado

Como ocurre siempre que sale a la luz un nuevo indicador suelen proliferar los estudios que analizan las correlaciones con otras métricas. El caso de las Library Holdings Counts o inclusions no ha sido una excepción y a la luz de los resultados se puede afirmar que ofrecen una visión diferente a los indicadores de citación. Aunque los estudios que han abordado la relación entre citas y libcitations lo han hecho casi todos empleando metodologías, fuentes de información y disciplinas diferentes sí se observa un patrón en todos: las correlaciones, aunque en ocasiones pueden llegar a ser significativas, suelen ser bajas y poco relevantes como por ejemplo ocurre en los estudios de Linmans (2010) y Zuccala y Guns (2013). Todo indica que las citas y las libcitations capturan cierta información en común pero que no existe una relación causa-efecto en ninguna de las direcciones. Estamos ante un indicador que mide o captura un tipo de impacto o difusión diferente al de la citación.

Entre los estudios que confirman estos hechos (Tabla 3) el primero que encontramos es el de Linmans (2010) que abordó las correlaciones entre los library holdings con citas calculadas a partir de Web of Science en el contexto de la Facultad de Humanidades de la Universidad de Leiden. Las correlaciones que obtuvo en su conjunto fueron de 0.29, sin embargo si los libros estaban escritos en inglés aumentaba a 0.49. En otro estudio Zuccala & White (2015) ofrecen datos para dos disciplinas, dos períodos cronológicos y hacen distinción de si las bibliotecas pertenecían o no a la *Association of Research Libraries* (ARL). Zuccala y White realizan un análisis considerando el periodo de años 1996-2011, la base de datos Scopus y 10 disciplinas humanísticas. Para las dos principales disciplinas, la Historia y la Literatura, obtienen respectivamente una correlación de 0.24 y 0.20, estos valores se incrementan al seleccionar solo bibliotecas de la ARL llegando a 0.26 y 0.24. En líneas generales al considerar las diez disciplinas y dos períodos cronológicos es muy difícil que se sobrepase una correlación del 0.20 y en casos muy excepcionales alcanza el 0.28.

**Table 3. Correlaciones encontradas en diversos estudios entre el número de inclusiones en library holdings o libcitations y el número de citas**

| Author | Tipo de estudio | Citation database | Coeficiente de correlación |
|---|---|---|---|
| Linmans (2010) | Libros publicados en por una Facultad de Humanidades | Web of Science | 0.29 All books<br>0.40 Book in english |
| Kousha & Thewall (2016) | 759 libros libros publicados en Ciencias Sociales y 1262 Libros publicados en Humanidades | Book Citation Index | 0.145 Ciencias Sociales<br>0.141 Humanidades |
| Kousha & Thewall (2016) | 759 libros libros publicados en Ciencias Sociales y 1262 Libros publicados en Humanidades | Google Books | 0.234 Ciencias Sociales<br>0.268 Humanidades |
| Zuccala & White (2015) | 20 996 libros de historia y 7541 libros de Literature & Literary Theory citados en revistas Scopus durante el período 2007-2011 | Scopus | 0.24 History<br>0.20 Literature & Literary Theory |
| Zhang, Zhou, & Zhang (2018) | 2356 indexados en el Chinese Social Sciences Citation Index | Chinese Social Sciences Citation Index | 0.291 Ethnology<br>< 0.20 Resto disciplinas |

Con otras fuentes de información como el *Book Citation Index* y *Google Books* Kousha y Thewall (2015) encuentran las correlaciones más altas en este último, alcanzando el 0.234 en Ciencias Sociales y 0.268 en el caso de las Humanidades. En otros contextos lingüísticos tampoco se han encontrado correlaciones especialmente positivas. Zhang, Zhou y Zhang (2018) llevaron a cabo la correlación de las citas de 2356 libros incluidos en el *Chinese Social Sciences Citation Index* con sus respectivas inclusiones en library holdings (OCLC) en 21 disciplinas científicas y de todas ellas alcanza el valor más alto Ethnology con un coeficiente de correlación de Spearman de 0.291, el resto de las áreas se sitúan sistemáticamente por debajo de 0.2.

**Gráfico 1. Coeficiente de correlación de Spearman de los diversos indicadores incluidos en *PlumX Analytics* con el indicados de Library Holdings Counts.**

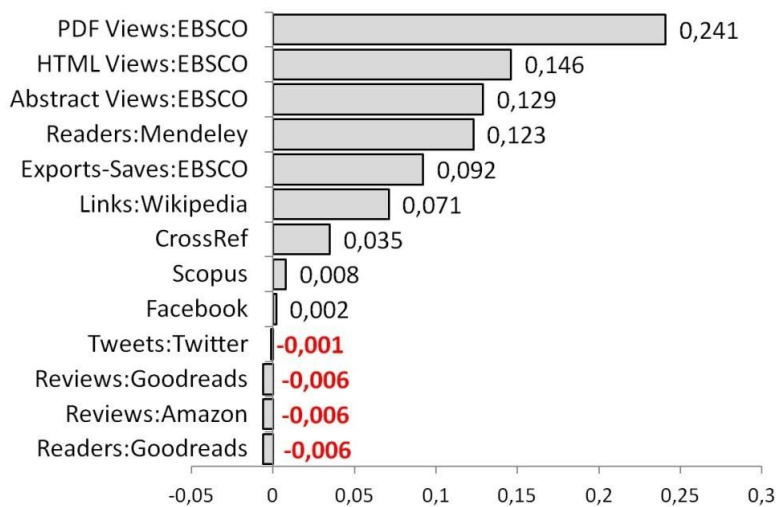

Nota: gráfico elaborado a partir de los datos incluidos en los materiales complementarios incluidos en el análisis de *PlumX Analytics* realizado por Torres-Salinas, Gumpenberger y Gorraiz (2017)

Otros estudios han abordado la relación de las Library Holdings con las altmétricas. Kousha y Thewall (2016) analizaron la correlación de las libcitacions de un conjunto de libros indexados en el *Books Citation Index* con diversos indicadores de Amazon (reviews, stars, sales rank, …). Tanto para el caso de las Ciencias Sociales (n=759) como de las Humanidades (n=1262) la mayor correlación se obtuvo con el indicador Amazon Reviews con 0.321 y 0.348 respectivamente. En el caso de las Ingenierías (n=718) todas las correlaciones eran sustancialmente menores como demuestra su valor más elevado de 0.129 con las reviews. También Torres-Salinas, Gumpenberger y Gorraiz (2017) muestran la baja correlación de las library counts con diferentes altmetricas incluidas en *PlumX Analytics* (Gráfico 1). En este caso la muestra era un colección de 268 061 libros y las correlaciones más altas las tenía con tres indicadores relacionados con las vistas de la plataforma EBSCO (vistas PDF=0.241, html=0.146, abstract=0.129).

A la luz de los resultados está claro que nos encontramos ante un indicador que mide una dimensión diferente a la del impacto científico. Kousha y Thewall (2016) apuntan a la capacidad de las Libcitations para capturar el valor educativo[22] de los libros y su influencia cultural. Otros autores ofrecen una visión simplificada del asunto indicando que la presencia en bibliotecas refleja la recepción positiva o negativa de un libro (Biagetti, 2018b) pero que en cualquier caso no es una medida de evaluación científica. Para nosotros el indicador de Library Inlusions o Counts, cuando las bibliotecas empleadas son académicas, es una expresión de la potencial utilidad de un libro para la comunidad científica (Torres-Salinas, 2009). Esta visión también la mantienen Zuccala y White (2015) afirmando por un lado que las inclusiones en bibliotecas reflejan "the potential readerships, or users, of a given book" y por otro lado que no son reflejo del impacto social, si bien los libros que han producido cambios sociales presenta alto este indicador (White et al. 2009; Zuccala & White 2015). En cualquier caso el significado cambia dependiendo de la muestra de bibliotecas que se utilice. Si solo analizamos las indexaciones que tienen una colección de libros en bibliotecas públicas el signifcado de las libcitations puede aproximarse más al impacto cultural, pero si analizamos bibliotecas académicas podría aproximarse más a la utilidad y el impacto educativo.

Otros aspectos que podrían cuestionar el valor y significado de estos indicadores es que no siempre los libros presentes en un catálogo responden a una selección de un bibliotecario ya que muchos de ellos son donaciones, regalos, etc. Desde esta perspectiva el más crítico es Biagetti (2018b) que señala que los libros no siempre se escogen por elección consciente y habla de lo determinadas que están muchas de las compras a través de las adquisiciones cooperativas, los planes de compras o las peticiones de particulares. Las donaciones según Biagetti (2018a) son uno de los puntos que más crítica e indica que el 4.2% de los libros (n=2165) en Bibliography and Library Science y el 8.2% (n=1097) en History of Political Institutions eran donaciones. A la luz de los resultados de Biagetti no se deberían considerar las donaciones como un problema relevante y más bien habría que tratarlo como un factor similar al de la autocitación o la cita negativa. Críticas similares a las anteriores las encontramos también en Hammarfelt (2016, p. 122) que apunta a las compras en piloto automático que realizan las bibliotecas y que no implican mérito para el autor. Pero quizás el factor más distorsionante en el valor de la libcitations, como fruto de un proceso

---

[22] Véase al respecto la Tabla 8 de este trabajo, donde se confirma o se percibe esta idea del valor educativo de los libros, una dimensión no captada por la citación, ya que los manuales o los handbooks no suelen ser citados en la literatura científica.

selectivo, sea la compra de e-books ya que éstos se integran de forma masiva en las colecciones de las bibliotecas a través de paquetes y bases de datos (Lewis & Kennedy, 2019)

En cualquier caso, al margen de las dudas que se puedan producir por el proceso de adquisición, consideramos que los Library Holdings son capaces de captar la utilidad potencial de un libro. Los libros se adquieren por parte de las bibliotecas especializadas para ser utilizados por sus usuarios (estudiantes e investigadores). Esta adquisición implica un juicio valorativo que suele ser resultado de un Plan Selection o bien resultado de un Patron-Driven Acquisitions (PDA) (Tyler et al., 2019), es decir, una adquisición que se realiza en función de los usos reales que haya tenido (Yuan, Van Ballegooie, & Robertson, 2018). Cuando un libro está en muchas bibliotecas implica al menos dos cosas: 1) se ha percibido la utilidad de un libro bien sea por parte de un bibliotecario o bien por parte de un usuario (PDA o librarians orders) y 2) el libro al estar en muchas bibliotecas tiene un público y una utilidad latente mayor que aquellos que no lo están. Asimismo el tipo de utilidad (de carácter cultural, científica o educativa) viene determinada por el tipo de biblioteca que utilicemos como fuente de información. En cualquier caso hay que indicar que el Library Holding Inclusions es un indicador que no debería utilizarse de manera aislada para determinar el impacto científico, tal y como ocurre con el resto altmétricas (Haustein, 2016).

## 4. Fuentes de información

En primer lugar hay que considerar que el LCA no tiene una metodología uniforme y que cada análisis conlleva el diseño de una metodología específica. Como hemos comentado anteriormente, el factor más importante es la selección de las bibliotecas, que determinará los resultados y su interpretación. Por ello a la hora de afrontar un estudio de library holdings tenemos que considerar tres aspectos relacionados con las fuentes de información:

a) Qué bibliotecas vamos a consultar
b) Qué fuentes de información vamos a emplear
c) Cómo vamos a recuperar la información

En relación a las bibliotecas la mayor parte de los estudios acotan la búsqueda de libros a un conjunto concreto de OPACS. El uso selectivo de determinados catálogos en función de ciertas características (prestigio institucional, ámbito lingüístico, etc…) es una cuestión que ya se manifestó en los primeros estudios. Torres-Salinas y Moed (2009), por ejemplo, seleccionaron solamente 42 bibliotecas de universidades de prestigio en Economía. Sin embargo si afrontamos un análisis en un contexto no anglosajón (España, Italia, China …) o en disciplinas con contenidos con un carácter muy local (Arqueología, Historia, etc..) es conveniente seleccionar catálogos del propio país o del contexto geográfico. Así lo hace Biagetti (2018a) que configura una muestra mixta balanceada de catálogos de todo el mundo pero incluyendo 13 bibliotecas italianas. Por su parte, Linmans (2010) selecciona bibliotecas tan solo de tres países Estados Unidos, Reino Unido y Países Bajos. Asimismo al hacer uso de catálogos colectivos genéricos es conveniente eliminar determinadas tipologías de bibliotecas si lo que se busca es analizar el impacto científico, por ello Zuccala y White (2015) discriminan las Libcitations en función de la biblioteca de procedencia, en su caso si provienen de miembros de la *Association of Research Libraries* (ARL) o de no miembros. Justifican los autores esta discriminación señalando que las bibliotecas públicas no tienen como objetivo primario contribuir y ayudar a la ciencia.

Desde el punto de vista de las fuentes de información primarias son habituales desde hace años los catálogos colectivos de bibliotecas, que permiten lanzar consultas a diversas bibliotecas desde una plataforma única gracias al protocolo Z39.50. Es habitual que se creen a nivel nacional catálogos colectivos de bibliotecas universitarias y especializadas como es el caso de *Jisc Library Hub Discover* en Reino Unido[23] e Irlanda, *SUDOC*[24] en Francia, *REBIUN*[25] en España, *INKA* en Alemania[26] o *LIBIS* en Bélgica[27]. Estos OPACS identifican en qué bibliotecas estaría indexado un libro y son fuentes útiles para estudios en pequeña escala en contextos nacionales o locales. Sin embargo, la importancia y uso de los catálogos colectivos nacionales ha quedado mitigada por WorldCat de la OCLC que tiene capacidad para aglutinar miles de catálogos a nivel mundial por lo que no es de extrañar que se haya convertido en la base de datos por excelencia y la que ha prevalecido para la realización de los estudios de catálogos de bibliotecas (Linmans, 2010; Zuccala & White, 2015; Neville & Henry, 2014; Halevi, Nicolas, & Bar-Ilan, 2016).

**Table 4. Users location and number of academic libraries in Worldcat**

| Countries | Users Location[note1] | Number of academic libraries[note2] | Number of public libraries[note2] | Number of other type of libraries[note2] | Total number of libraries[note2] |
|---|---|---|---|---|---|
| United States | 44.8% | 2505 (43.16%) | 3532 (79.53%) | 3478 (70.26%) | 9515 (62.62%) |
| United Kingdom | 3.7% | 137 (2.36%) | 131 (2.95%) | 88 (1.78%) | 356 (2.34%) |
| Germany | 3.2% | 307 (5.29%) | 18 (0.41%) | 89 (1.8%) | 414 (2.72%) |
| France | 2.3% | 1113 (19.18%) | 9 (0.2%) | 34 (0.69%) | 1156 (7.61%) |
| Italy | 1.7% | 104 (1.79%) | 113 (2.54%) | 20 (0.4%) | 237 (1.56%) |
| Spain | 1.5% | 42 (0.72%) | 6 (0.14%) | 19 (0.38%) | 67 (0.44%) |

[note1] Geographical location of users, results from log (Wakeling et al., 2017)
[note2] Data from the Directory of OCLC members

Una de la razones del éxito de WorldCat es su tamaño, para hacernos una idea del mismo en la actualidad en el directorio de la OCLC se pueden identificar un total 15 194 bibliotecas de las cuales 5804 son académicas[28]. Sin embargo pese a la evidente ventaja que supone WorldCat pocos estudios han abordado críticamente la utilización de este producto sobre todo considerando el claro sesgo anglosajón que presentan (Wakeling et al., 2017). En la Tabla 4 se señala que el 44,8% de los usuarios de esta plataforma y el 43% de las bibliotecas académicas son estadounidenses, cifras muy superiores a las de cualquier país europeo. Por ejemplo solo el 1,5% y el 0,7% de la bibliotecas son españolas lo que sin duda nos obliga a utilizar fuentes complementarias al analizar dentro de España la difusión de libros, ya que de 76 bibliotecas universitarias tan solo 42 están representadas en WorldCat.

---

[23] Jisc Library Hub Discover: https://discover.libraryhub.jisc.ac.uk/
[24] SUDOC - Catalogue du Système Universitaire de Documentation: http://www.sudoc.abes.fr/xslt/
[25] REBIUN - Red de Bibliotecas Universitarias Españolas: https://www.rebiun.org
[26] INKA - Inkunabelkatalog: http://www.inka.uni-tuebingen.de
[27] LIBISnet: http://libis.be/libis/libisnet
[28] Información extraída del Directoriy of OCLC Members: https://www.oclc.org/en/contacts/libraries.html. Hay que indicar que algunos lugares de su propia la OCLC eleva el número de sus bibliotecas miembro a 17,983: https://www.oclc.org/en/about.html

Es recomendable por ello, siempre que llevemos a cabo un LCA con WorldCat, consultar el directorio de miembros de la OCLC para verificar la distribución territorial de los catálogos.

La tercera cuestión a la que debemos atender es cómo vamos a consultar las bibliotecas y recuperar la información. La primera forma es consultar los catálogos a través de su buscador genérico pero esta opción implica ejecutar consultas manuales o bien aplicar alguna técnica de web scrapping. Otra de las opciones para la recuperación de datos bibliográficos es mediante un cliente Z39.50 entre el que podemos destacar *Bookwhere Suite de WebClarity*[29]. Sin embargo solo Torres-Salinas y Moed (2008) hacen uso de esta metodología, que pese a permitir crear grandes bases de datos con un alto grado de normalización también conlleva diversos problemas: configuración manual del Z39.50 de cada biblioteca, bibliotecas que no comparten sus datos de conexión públicamente, errores de conexión, etc... Por ello este método apenas se ha utilizado y lo más habitual en este tipo de estudios es emplear las propias APIs de los diferentes catálogos.

En relación a las APIS muchas bibliotecas individuales lo tienen, como Harvard Library[30] al igual que catálogos colectivos como por ejemplo COPAC[31]. También WorldCat cuenta con una API[32], que es útil para este tipo de estudios ya que se puede lanzar una búsqueda usando diversos parámetros (ISBN, OCLC number, …) y obtener las bibliotecas y sus localizaciones. Los resultados los devuelve la API en formato MARC XML o Dublin Core formats para los registros bibliográficos y en XML o JSON para los library catalog URL y library locations (Tabla 5), siendo el límite de consultas de 50 000 al día. Como contrapartida, el acceso es otorgado a una institución miembro que tutela y es responsable del uso[33], lo que no la hace totalmente libre e independiente para el mundo de la investigación.

**Tabla 5. Supported Operations para los library locations en la API de WorldCat**

| Library Locations | Información que ofrece WorldCat | URL path |
| --- | --- | --- |
| GetByOCLCNumber | Get Library Locations by OCLC Number and geographic information | /content/libraries/{OCLC_Number} |
| GetByISBN | Get library locations by ISBN near a geographic location | /content/libraries/isbn/{ISBN} |
| GetByISSN | Get library locations by ISSN near a geographic location | /content/libraries/issn/{ISSN} |
| GetByStandardNumber | Get library locations by Standard Number and geographic location | /content/libraries/sn/{Standard_Number} |

---

[29] Página web: http://www.webclarity.info/products/bookwhere/
[30] Harvard Library APIs & Datasets: https://library.harvard.edu/services-tools/harvard-library-apis-datasets
[31] COPAC API: https://www.programmableweb.com/api/copac
[32] WorldCat Search API: https://platform.worldcat.org/api-explorer/apis/wcapi
[33] WorldCat Search API for OCLC Member Institution: Terms and Conditions:
https://www.oclc.org/content/dam/developer-network/PDFs/wcapi-terms-and-conditions-20121204.pdf

Por otro lado las fuentes que se han impuesto en el mundo de las altmétricas son los agregadores capaces de recuperar un conjunto indicadores para un set de documentos con solo ingresar algún tipo de número normalizado (DOI, Handle, etc.). Entre estos agregadores, por la especial atención que le ha prestado a las altmeétricas de libros y monografías, destaca *PLumX Analytics* (Torres-Salinas, Gumpenberger & Gorraiz, 2017) ya que, por ejemplo, estudios recientes demuestran la poca eficiencia de *Altmetric.com* a la hora de valorar el libro (Torres-Salinas, Gorraiz, & Robinson-García, 2018). PlumX, actualmente propiedad de Elsevier, incluye entre sus indicadores el Library Holdings Inclusions en WorldCat (Holdings: WorldCat) y nos permite fácilmente su cálculo a partir de los ISBN y consultas masivas. Uno de los primeros trabajos en utilizar esta fuente es el de Halevi, Nicolas y Bar-Ilan (2016) que emplea 71 443 eBooks ISBNs numbers de la the Levy Library Ebrary collection para realizar un ranking con los libros con más inclusiones de los cuales 98.80% estaban indexados en Worldcat.

En consonancia con los estudios anteriores Torres-Salinas, Gumpenberger y Gorraiz (2017a) también trabajaron con PlumX con una colección de 263 210 libros, de los cuales el 97.81% estaban incluidos en alguna biblioteca, 18 indicadores fueron analizados y WorldCat en PlumX tuvo la mejor cobertura y los valores más elevados (Tabla 6). Otros trabajos confirman el valor de PLumX, como el de Torres-Salinas, Robinson-García y Gorraiz (2017b). Este estudio es relevante ya que analiza una muestra (n=2957) de libros publicados por investigadores de la Universidad de Granada y descubriendo que 48% de los hits métricos para libros correspondían a Library Holdings Inclusions y el 79% de los libros presentaban esta métrica. El estudio evidencia por un lado el acierto de PlumX al incluir indicadores basados en WorldCat y la potencia de los library holdings counts frente a otros indicadores altmétricos.

**Tabla 6. Cobertura de diferentes indicadores incluidos en el agregador PLumX a partir de diferentes estudios.**

| Indicators | Results from Halevi* | Results from Torres-Salinas** |
|---|---|---|
| Holdings WorldCat | 98.80% | 97.81% |
| Abstract views: EBSCO | 91.52% | 95.12% |
| Saves: EBSCO | 78.19% | 81.77% |
| PDF views: EBSCO | 65.64% | 68.28% |
| Goodreads captures | 69.23% | 53.70% |
| Mendeley captures | 43.11% | 24.86% |
| Goodreads reviews | 25.08% | 19.13% |
| Wikipedia links | 25.47% | 16.57% |
| Citations Scopus or CrossRef | 4.04% | 4.25% |
| *Halevi: 2016. n=71 443. Levy Library Ebrary collection <br>**Torres-Salinas 2017. n=263 210. University of Vienna, the EBSCO e-book collection | | |

Como consideraciones metodológicas finales hay que reseñar que a la hora de llevar a cabo un LCA debemos tener en cuenta las características formales propias de los libros, sobre todo cuando se emplean métodos masivos de recopilación de datos, como los títulos con diferentes ISBN por tener varios formatos o ediciones, las posibles traducciones que se hayan realizado a otros idiomas o las diferencias de catalogación de los distintos países (Biagetii, 2018a). Todas estas cuestiones, ya reseñadas en los primeros estudios (Linmans, 2010), provocan problemas en el matching de títulos e ISBN afectando sobre todo a los catálogos colectivos que integran información muy heterogénea; estos problemas se reflejan por ejemplo en las diferencias de conteo de bibliotecas entre la página principal de WorldCat, la API o WorldCat Identities.

## 5. Experimento usando WorldCat Identities

Uno de los niveles de agregación en el que todavía no hemos visto aplicado los Library Holdings Counts es a nivel de autor; asimismo y pese a sus posibilidades tampoco ha sido empleada en ningún estudio *WorldCat Identities*. En este apartado nos aproximamos experimentalmente y con un nuevo enfoque a la aplicación del LCA a nivel de autor y disciplinar, en este caso la informetría, utilizando la mencionada WorldCat Identities. Esta plataforma unifica bajo un mismo encabezamiento normalizado, gracias a diversas técnicas de data mining y clustering[34], todos los "works" de un autor calculando los datos de la difusión en bibliotecas de su obra, tanto a nivel de autor como desglosado por publicación e integrando además otros datos de carácter contextual (géneros, topics, variantes del nombre, co-autores, etc…). Véase por ejemplo la Ilustración 1 donde se ofrece la información agrupada para Henk F. Moed.

**Ilustración 1. Información básica ofrecida por WorldCat Identities para un autor**

Moed, H. F.
Overview
Works: 45 works in 165 publications in 5 languages and 2,385 library holdings
Genres: Handbooks and manuals  Conference papers and proceedings
Roles: Author, Editor, Other, Creator
Classifications: PN171.F56, 001.42

Pues bien tomando como fuente WorldCat Identities se ha consultado una muestra representativa de 22 investigadores de prestigio del ámbito de la informetría. Hemos recopilado para cada uno de ellos la información agrupada de WorldCat Identities sobre Library Holdings, Publicaciones/Works que a su vez hemos completado con el número de citas en los perfiles de Google Scholar. Los resultados se muestran en la Tabla 7. El autor con mayor cantidad de Library Holdings Counts para el conjunto de sus obras es Blaise Cronin con un total 6749, seguido de Chaomei Chen y Leo Egghe que suman respectivamente 5867 y 3718. En el caso de Henk Moed ocupa el quinto lugar de esta clasificación con 2385.

No siempre ocupan los primeros puestos de la clasificación según Libcitations los autores más citados. Por ejemplo en esta situación se encuentran Loet Leydesdorff, Mike Thelwall o más claramente Ton van Raan (406 bibliotecas; 13857 citas). Evidentemente este listado no está basado en artículos científicos sino en monografías, conferencias u obras donde el investigador aparece como contributors de capítulos. Por tanto en el ranking de Library Holdings están especialmente representados todos aquellos que han escrito o editado handbooks o manuales. Esta clasificación reflejaría un aspecto basado en otro tipo de actividades y contribuciones académicas, relacionadas con la generación de contenidos docentes/educativos (por ejemplo manuales y libros profesionales) o el engagement del autor dentro de su campo (por ejemplo la edición como editor de unas actas). Existen autores que claramente, no solo han contribuido científicamente a un campo sino que además han desarrollado otras actividades o han publicado otro tipo de materiales más allá de los artículos. Sírvanos como ejemplo de este perfil Chaomei Chen o el propio Henk Moed. La clasificación, como ya afirmábamos en el apartado sobre el signifcado, capta un valor de la actividad académica que queda fuera de la citación.

---

[34] WorldCat Identities cuenta con 30 millones de entradas, agrupa diferentes fuentes de información como VIAF y FAST. Másc información https://www.oclc.org/research/themes/data-science/identities.html

Hemos querido completar este perfil de la informetría con Library Holdings Counts a través de WorldCat Identities con un listado de los libros más destacados que nos ayudan a completar la Tabla 7 lo que nos has permitido distinguir dos fenómenos antagónicos. En primer lugar hay una serie de libros con gran impacto científico y difusión en bibliotecas; con este perfil el título que más destaca es el libro de Henk F. Moed "*Citation analysis in research evaluation*" que es el más citado (1711 citas en Google Scholar) y el tercero indexado en mayor número de bibliotecas, 1010. Otros libros con un perfil similar serían el de Chen "*Mapping scientific frontiers: the quest for knowledge visualization*" (808 bibliotecas; 401 citas) o el de De Bellis "*Bibliometrics and citation analysis: from the Science Citation Index to cybermetrics*" (509 bibliotecas; 728 citas). Nos encontramos ante libros que tienen un gran reconocimiento científico y que además son manuales o handbooks de referencia, un valor de este último que es capaz de captar el indicador de library counts.

**Tabla 7. Listado de investigadores del ámbito de la informetría ordenados según el número Library Holdings de WorldCat Identities.**

| Autor Worldcat Entities | Works y publications en Worldcat Entities[35] | Número de Library Holdings | Citas deGoogle Scholar |
|---|---|---|---|
| Cronin, Blaise | 144 works in 582 publications | 6749[36] | 11 122 |
| Chen, Chaomei | 42 works in 243 publications | 5867 | 15 968 |
| Egghe, L. (Leo) | 57 works in 186 publications | 3718 | -- |
| Garfield, Eugene | 150 works in 447 publications | 3386 | 30 105 |
| Moed, H. F. | 45 Work, 165 Publications | 2385 | 13 026 |
| Sugimoto, Cassidy R. | 10 works in 85 publications | 2270 | 5894 |
| Braun, Tibor | 156 works, 389 publications | 2268 | 9430 |
| Wolfram, Dietmar | 15 works in 49 publications | 1769 | -- |
| Debackere, Koenraad | 105 works, 175 publications | 1628 | 9762 |
| Ingwersen, Peter | 33 works in 142 publications | 1608 | 11 316 |
| Rousseau, R. | 25 works in 121 publications | 1385 | 13 534 |
| Rowlands, Ian | 22 works in 92 publications | 1298 | 5188 |
| Leydesdorff, L. A. | 64 works, 189 publications | 1230 | 47 889 |
| Thelwall, Mike | 46 works in 113 publications | 1132 | 28 585 |
| Glänzel, Wolfgang | 53 works, 114 publications | 1115 | 18 238 |
| De Bellis, Nicola | 7 works, 25 publications | 762 | -- |
| Narin, Francis | 45 works, 96 publications | 426 | 15 324 |
| Raan, A. F. J. van | 32 works, 68 publications | 406 | 13 857 |
| Schubert, András | 21 works, 62 publications | 394 | 12 816 |
| Persson, Olle | 121 works,174 publications | 257 | 5205 |
| Bornmann, Lutz | 14 works, 28 publications | 215 | 14 945 |
| Nederhof, A. J. | 38 works, 59 publications | 199 | -- |

Un segundo grupo, más controvertido, son los libros con mucha presencia en bibliotecas, pero poco citados; ocurre por ejemplo con el libro de Holmberg "*Altmetrics for information professionals: past, present and future*" (745 bibliotecas; 41 citas) o el de Ding et al. "*Measuring scholarly impact : methods and practice*" (428 bibliotecas; 2 citas). Son libros con una claro perfil profesional y práctico no orientados a un público científico, y que gracias a las library counts pueden analizarse desde otro prisma. Independientemente del orden el listado ofrecido en la Tabla 8 no es ajena a ningún especialista de la informetría y sería difícil negar que todos ellos son referencia en el área, ya sea desde un punto de vista científico o bien docente/profesional.

Finalmente, y considerando que este libro es un homenaje a Henk F. Moed hay que subrayar su contribución al campo ya que consigue situar tres libros en este listado de las obras sobre informetría más populares en bibliotecas. Junto al ya mencionado "*Citation

---

[35] El indicador publicación hay que tomarlo con precaución ya que una misma obra puede tener varias ediciones, entendidas estas últimas por ejemplo en función del número de ediciones, el tipo de formato de la obra (print o ebook) o incluso una misma obra puede estar catalogada en un biblioteca de forma diferente a como lo está en otra, en este mismo caso WorldCat las entiende como "ediciones" diferentes.

[36] Para el cálculo de este indicador se incluyen las inclusiones de los libros homenaje (Festschrift). Por ejemplo en el caso Cronin se incluyen 921 inclusiones que recibe el libro Theories of informetrics and scholarly communication: a Festschrift in honor of Blaise Cronin

*analysis in research evaluation*" encontramos también el "*Handbook of quantitative science and technology research*" (832 bibliotecas) del cual fue editor y su última monografía "*Applied evaluative informetric*" (298 bibliotecas). Si por algo ha destacado Henk ha sido no solo por contribuciones al desarrollo de la disciplina (Crown Indicator, SNIP, etc…) sino también, especialmente en la última etapa de su vida académica, de hacer más accesible la bibliometría a círculos profesionales con manuales que ya podemos considerar clásicos y de referencia.

**Tabla 8. Libros sobre informetría con mayor difusión en bibliotecas miembro de WorldCat**

| Referencia Bibliográfica[37] | Bibliotecas miembro de WorldCat | Citas de Google Scholar |
|---|---|---|
| **Egghe, Leo**. *Power laws in the information production process : Lotkaian informetrics*. Amsterdam: Elsevier/Academic Press, 2005. | 1255 | 322 |
| **Wolfram, Dietmar**. *Applied informetrics for information retrieval research*. Westport, Conn. : Libraries Unlimited, 2003. | 1166 | 76 |
| **Moed, Henk**. *Citation analysis in research evaluation.* Dordrecht : Springer, 2005. | 1010 | 1711 |
| **Sugimoto, Cassidy R** (editor). *Theories of informetrics and scholarly communication: a Festschrift in honor of Blaise Cronin*. Berlin : De Gruyter, 2016 | 921 | 11 |
| **Moed, Henk; Glänzel, Wolfgang; Schmoch, Ulrich** (editors). *Handbook of quantitative science and technology research : the use of publication and patent statistics in studies of S & T systems*. Dordrecht Springer, 2011 | 832 | 328 |
| **Chaomei, Chen**. *Mapping scientific frontiers : the quest for knowledge visualization*. London: Springer, 2013 | 808 | 441 |
| **Holmberg, Kim**. *Altmetrics for information professionals: past, present and future*. Waltham: Chandos, 2016 | 745 | 41 |
| **Gardfield, Eugene**. *Citation indexing - its theory and application in science, technology, and humanities*. Philadelphia: ISI Press, 1983 | 686 | 2924 |
| **Chaomei, Chen**. *CiteSpace : a practical guide for mapping scientific literature*. New York : Nova Science Publishers, Inc., 2016 | 571 | 36 |
| **Cronin, Blaise**. *The hand of science : academic writing and its rewards*. Lanham, Md : Scarecrow Press, 2005. | 515 | 267 |
| **De Bellis, Nicola**. *Bibliometrics and citation analysis : from the Science Citation Index to cybermetrics*. Lanham, Md. : Scarecrow Press, 2009. | 509 | 728 |
| **Cronin, Blaise**. *Beyond bibliometrics : harnessing multidimensional indicators of scholarly impact*. Cambridge : The MIT Press, 2014 | 502 | 121 |
| **Ding, Ying; Wolfram, Dietmar; Rousseau, Ronald** (editors). *Measuring scholarly impact : methods and practice*. Cham : Springer, 2014 | 428 | 2 |
| **Leydesdorf, ,Loet; Besselaar, Peter Van Den** (editors). *Evolutionary economics and chaos theory : new directions in technology studies*. New York : St. Martin's Press, 1994. | 331 | 199 |
| **Thewall, Mike**. *Introduction to webometrics : quantitative web research for the social sciences*. San Rafael, Calif : Morgan & Claypool Publishers, 2009. | 328 | 375 |
| **Leydesdorf, ,Loet**. *Universities and the global knowledge economy : a triple helix of university-industry-government relations*. Pinter Pub Ltd 2005. | 306 | 67 |
| **Moed, Henk**. *Applied evaluative informetric*. Springer, 2017 | 298 | 43 |
| **Anne-Wil Harzing**. *The publish or perish book : your guide to effective and responsible citation analysis*. Melbourne, Australia : Tarma Software Research Pty Ltd, 2013. | 306 | 316 |
| **Vaan Raan, Anthony** (editor). *Handbook of quantitative studies of science and technology*. Amsterdam : Elsevier Science, 2014. | 232 | 273 |

---

[37] Para la realización de este listado se han considerado las obras con mayor library counts de los autores incluidos en la Tabla 8. Asimismo se han realizado búsquedas mediante palabras clave como informetrics, bibliometrics, altmetrics, citation analysis, lo que ha permitido identificar otras obras como la de Kim Holmberg, Anne-Wil Harzing o Nicola De Bellis.

## 6. Consideraciones finales

En este capítulo hemos querido hablar de diversos aspectos del Library Catalog Analysis como homenaje a Henk F. Moed que contribuyó intelectualmente a su conceptualización en uno de los artículos seminales. En este sentido Moed apoyó y vislumbró la utilidad de estas nuevas métricas en un momento donde existía un déficit de indicadores aplicados a los libros. Actualmente los library holdings, gracias sobre todo a WorldCat y a PlumX, forman el set de las altmétricas y tal y como han evidenciado diversos estudios es uno de los indicadores que mejor cobertura ofrecen frente a otros como las reseñas en Goodreads o Amazon. De esta forma las Library Holdings Inclusión se perfilan como un complemento ideal para combinar con la citación. Como hemos comprobado en el ámbito de la informetría reflejan una utilidad profesional o educativa siendo especialmente útil para analizar aquellas monografías no orientadas a públicos científicos.

## 7. Bibliografía

Archambault, É., Vignola-Gagné, É., Côté, G., Larivière, V., & Gingrasb, Y. (2006). Benchmarking scientific output in the social sciences and humanities: The limits of existing databases. *Scientometrics*, *68*(3), 329–342. https://doi.org/10.1007/s11192-006-0115-z

Biagetti, M. T., Iacono, A., & Trombone, A. (2018a). Is the Diffusion of Books in Library Holdings a Reliable Indicator in Research Assessment? BT - The Evaluation of Research in Social Sciences and Humanities: Lessons from the Italian Experience. In A. Bonaccorsi (Ed.) (pp. 321–343). Cham: Springer International Publishing. https://doi.org/10.1007/978-3-319-68554-0_14

Biagetti, M. T., Iacono, A., & Trombone, A. (2018b). Testing library catalog analysis as a bibliometric indicator for research evaluation in Social Sciences and Humanities. In *Challenges and Opportunities for Knowledge Organization in the Digital Age: Proceedings of the Fifteenth International ISKO Conference 9-11 July 2018 Porto, Portugal*(1st ed., pp. 892–899). Baden-Baden: Ergon-Verlag. https://doi.org/10.5771/9783956504211-892

Hammarfelt, B. (2016). Beyond Coverage: Toward a Bibliometrics for the Humanities BT - Research Assessment in the Humanities: Towards Criteria and Procedures. In M. Ochsner, S. E. Hug, & H.-D. Daniel (Eds.) (pp. 115–131). Cham: Springer International Publishing. https://doi.org/10.1007/978-3-319-29016-4_10

Haustein, S. (2016). Grand challenges in altmetrics: heterogeneity, data quality and dependencies. *Scientometrics*, *108*(1), 413–423. https://doi.org/10.1007/s11192-016-1910-9

Halevi, G., Nicolas, B., & Bar-Ilan, J. (2016). The Complexity of Measuring the Impact of Books. *Publishing Research Quarterly*, *32*(3), 187–200. https://doi.org/10.1007/s12109-016-9464-5

Hicks, D. (1999). The difficulty of achieving full coverage of international social science literature and the bibliometric consequences. *Scientometrics*, *44*(2), 193–215. https://doi.org/10.1007/BF02457380

Huang, M., & Chang, Y. (2008). Characteristics of research output in social sciences and humanities: From a research evaluation perspective. *Journal of the American*

Society for Information Science and Technology, 59(11), 1819–1828. https://doi.org/10.1002/asi.20885

Kousha, K., & Thelwall, M. (2015). Web indicators for research evaluation: Part 3: books and non standard outputs. *El Profesional de La Información*, *24*(6), 724–736. https://doi.org/10.3145/epi.2015.nov.04

Kousha, K., & Thelwall, M. (2016). Can Amazon.com reviews help to assess the wider impacts of books? *Journal of the Association for Information Science and Technology*, *67*(3), 566–581. https://doi.org/10.1002/asi.23404

Lewis, R. M., & Kennedy, M. R. (2019). The Big Picture: A Holistic View of E-book Acquisitions. *Library Resources & Technical Services*, *63*(2), 160. https://doi.org/10.5860/lrts.63n2.160

Linmans, A. J. M. (2008). *Een exploratieve studie van de onderzoeksprestaties van de Faculteit Letteren aan de Universiteit Leiden (in Dutch)*. Internal CWTS Report

Linmans, A. J. M. (2010). Why with bibliometrics the Humanities does not need to be the weakest link - Indicators for research evaluation based on citations, library holdings, and productivity measures. *Scientometrics*, *83*(2), 337–354. https://doi.org/10.1007/s11192-009-0088-9

Nederhof, A. J. (2006). Bibliometric monitoring of research performance in the social sciences and the humanities: A review. *Scientometrics*, *66*(1), 81–100. https://doi.org/10.1007/s11192-006-0007-2

Neville, T. M., & Henry, D. B. (2014). Evaluating Scholarly Book Publishers—A Case Study in the Field of Journalism. *The Journal of Academic Librarianship*, *40*(3), 379–387. https://doi.org/https://doi.org/10.1016/j.acalib.2014.05.005

Nilges, C. (2006). The Online Computer Library Center's Open WorldCat Program. Library Trends, 54(3), 430–447. https://doi.org/10.1353/lib.2006.0027

Priem, J., Taraborelli, D., Groth, P., & Neylon, C. (2010). Altmetrics: A manifesto. Retrieved from http://altmetrics.org/manifesto/

Torres-Salinas, D., Cabezas-Clavijo, Á., & Jiménez-Contreras, E. (2013). Altmetrics: New Indicators for Scientific Communication in Web 2.0. *Comunicar*, *21*(41), 53–60. https://doi.org/10.3916/C41-2013-05

Torres-Salinas, D., Gorraiz, J., & Robinson-Garcia, N. (2018). The insoluble problems of books: what does Altmetric.com have to offer? *Aslib Journal of Information Management*, *70*(6), 691–707. https://doi.org/10.1108/AJIM-06-2018-0152

Torres-Salinas, D., Gumpenberger, C., & Gorraiz, J. (2017a). PlumX As a Potential Tool to Assess the Macroscopic Multidimensional Impact of Books. *Frontiers in Research Metrics and Analytics*, *2*(July), 1–11. https://doi.org/10.3389/frma.2017.00005

Torres-Salinas, D., & Moed, H. (2008). Library catalog analysis is a useful tool in studies of social sciences and humanities. In *A New Challenge for the Combination of Quantitative and Qualitative Approaches. 10th International Conference on Science and Technology Indicators*. Viena.

Torres-Salinas, D., & Moed, H. F. (2009). Library Catalog Analysis as a tool in studies of social sciences and humanities: An exploratory study of published book titles in Economics. *Journal of Informetrics*, *3*(1), 9–26. https://doi.org/10.1016/j.joi.2008.10.002

Torres-Salinas, D., Robinson-Garcia, N., & Gorraiz, J. (2017). Filling the citation gap: measuring the multidimensional impact of the academic book at institutional level with PlumX. *Scientometrics*, *113*(3), 1371–1384. https://doi.org/10.1007/s11192-017-2539-z

Tyler, D. C., Hitt, B. D., Nterful, F. A., & Mettling, M. R. (2019). The Scholarly Impact of Books Acquired via Approval Plan Selection, Librarian Orders, and Patron-Driven Acquisitions as Measured by Citation Counts. *College & Research Libraries; Vol 80, No 4 (2019): May*. https://doi.org/10.5860/crl.80.4.525

Wakeling, S., Clough, P., Silipigni Connaway, L., Sen, B., & Tomás, D. (2017). Users and uses of a global union catalog: A mixed-methods study of WorldCat.org. *Journal of the Association for Information Science and Technology*, *68*(9), 2166–2181. https://doi.org/10.1002/asi.23708

White, H. D., Boell, S. K., Yu, H., Davis, M., Wilson, C. S., & Cole, F. T. H. (2009). Libcitations: A measure for comparative assessment of book publications in the humanities and social sciences. *Journal of the American Society for Information Science and Technology*, *60*(6), 1083–1096. https://doi.org/10.1002/asi.21045

White, H. D., & Zuccala, A. (2018). Libcitations, worldcat, cultural impact, and fame. *Journal of the Association for Information Science and Technology*, *69*(12), 1502–1512. https://doi.org/10.1002/asi.24064

Yuan, W., Van Ballegooie, M., & Robertson, J. L. (2018). Ebooks Versus Print Books: Format Preferences in an Academic Library. *Collection Management*, *43*(1), 28–48. https://doi.org/10.1080/01462679.2017.1365264

Zhang, H., Zhou, Q., & Zhang, C. (2018). Multi-discipline correlation analysis between citations and detailed features of library holdings. *Proceedings of the Association for Information Science and Technology*, *55*(1), 946–947. https://doi.org/10.1002/pra2.2018.14505501188

Zuccala, A., & Guns, R. (2013). Comparing book citations in humanities journals to library holdings: Scholarly use versus "perceived cultural benefit" (RIP). In *14th International Society of Scientometrics and Informetrics Conference, ISSI 2013*(Vol. 1, pp. 353–360). Amsterdam.

Zuccala, A., Guns, R., Cornacchia, R., & Bod, R. (2015). Can we rank scholarly book publishers? A bibliometric experiment with the field of history. *Journal of the Association for Information Science and Technology*, *66*(7), 1333–1347. https://doi.org/10.1002/asi.23267

Zuccala, A., & White, H. D. (2015). Correlating libcitations and citations in the humanities with WorldCat and scopus data. In S. A.A., S. A.A.A., S. C., A. U., & T. Y. (Eds.), *15th International Society of Scientometrics and Informetrics Conference, ISSI 2015*(pp. 305–316). Royal School of Library and Information Science, University of Copenhagen, Birketinget 6, Copenhagen S, DK-2300, Denmark: Bogazici Universitesi.